\documentstyle[12pt]{article}
\def\applt{\;\; {\lower3pt\hbox{$
{\buildrel < \over {\scriptstyle \sim} }$}}\;\;}
\def\baselinestrech{1.5}
\large
\begin{document}
\newcommand{\cl}{\centerline}
\newcommand{\beq}{\begin{equation}}
\newcommand{\eeq}{\end{equation}}
\newcommand{\beqa}{\begin{eqnarray}}
\newcommand{\eeqa}{\end{eqnarray}}
\def\d{{\rm d}}
%%%%%%%%%%%%%%%%%%%%%%%%
\begin{flushright} ANL-HEP-PR-94-18\end{flushright}
\vbox{\vskip 0.3 true in}
 \centerline{\large{{\bf Dispersive Methods and QCD Sum Rules}}}\par
\
\centerline{\large{{\bf for $\gamma$ $\gamma$ Collisions }}}\par

\vbox {\vskip 0.3 true in}

%\centerline{\bf , June 21, 1993}
\centerline{Claudio Corian\`{o}}
\centerline{High Energy Physics Division, Argonne National Laboratory}
\centerline{9700 South Cass, IL 60439, USA}
\vbox {\vskip 0.2 true in}

\centerline {ABSTRACT}
\medskip
\narrower{
It has been shown, in the case of meson photoproduction,
 that the power-law falloff
of these reactions can be described
by lowest order (real) sum rules, at moderate momentum transfer.
The phases of these processes, in this regime,
 are usually thought to be non-perturbative.
In a sum rule framework, however, they can possibly be described
by radiative corrections to the hadronic spectral densities
of the corresponding helicities, which become complex functions to order
$\alpha_s$,
and the effects of interference can be strongly enhanced by the presence
of the vacuum condensates in the dispersion relation.
It is shown that the imaginary parts of these complex corrections
have a factorized form and can be evaluated in a systematic fashion,
while their real parts, at the same perturbative order, are down by
at least 2 powers of momentum transfer.
The analysis is done at two loop level,
combining dimensional regularization and light-cone methods.
The calculations are performed for all the independent set
of scalar diagrams generated by the OPE.
The analytical bounds are identified and discussed.

%\vbox{\vskip 1.0 true in}
\noindent{\footnotesize email: coriano@hep.anl.gov}
\footnote{Work supported by the U.S. Department of Energy, Division of High
Energy Physics, Contract W-31-109-ENG-38}
\newpage

\section{Introduction}
Sum rules have been a valuable tool in the investigation of
the intermediate energy region of QCD.
More recently, it has been suggested that such methods can find
application in the context of elastic scatterings of Compton type \cite{CRS}.
Compared to sum rules for form factors,
which involve dispersion relations only at fixed $t$,
the description of elastic scattering has been
performed by sum rule based on dispersion
relations at both $s$ and $t$ fixed,
and with a limited domain of analyticity in the remaining variables.
 For scatterings at intermediate angles such new description of 2-photon
elastic processes compares favourably well
with standard factorization theorems \cite{coli}.
Such $extended$ sum rule methods can be helpful in
the attempt to understand on a firmer ground the
transition to perturbative QCD in simple elastic processes, and can have
potential application to a new entire class of direct reactions of Compton type
and to the corresponding crossed channels.

In fact, two-photon collisions are among the most interesting ways to study
these important aspects of elastic scattering at moderate momentum
and energy transfers.

Together with Li, we have analized in detail the main features of the
sum rules for pion Compton scattering and pion photoproduction \cite{H1H2}
and compared their predictions with those
derived from modified \cite{BS} \cite{LS}
 and standard factorization theorems \cite{BL}.

All the work presented so far  \cite{co} \cite{coli} for these reactions
is based on $real$ sum rules for the
two helicities $H_1$ and $H_2$.

The analysis of the radiative corrections to the lowest order
result are, however,
of crucial importance in the study of the phases of these
amplitudes, since nontrivial interference effects between those can
be generated by such corrections.
Differently from the usual prescriptions based on factorization
theorems, such effects, in the sum rule context,
can be strongly enhanced due to the
presence of the vacuum condensates as fundamental parameters
in the dispersion relations.

It is widely believed that in the intermediate energy region of QCD
($Q^2$ $\approx$ 4-7 $Gev^2$) such phases cannot be incorporated
in a direct perturbative treatment and are of non-perturbative origin.
In the sum rule approach
interference effects are generated by complex contributions
(due to gluonic exchange) to the
lowest order (real) spectral densities.

The evaluation of such corrections, however, is a notrivial task and requires
considerable effort.
In this work (section 3)
we discuss in great detail the properties of analyticity
of the lowest order spectral densities
of those correlators
which interpolate with processes of
Compton type \cite{CRS}.
The motivation of our study relies mainly on the fact that
the region of analyticity of 4-point correlators, compared
to that for the 3-point correlators which are commonly used
in the sum rules for vertex functions,
has unobvious features.
Such a region turns out to be bound in size
by $u$-channel singularities, a moving $u$-channel cut, whose
location varies
when the other variables which appear in the dispersion integral are varied.
The study of the properties of analyticity of the spectral densities
which appear in sum rules of these type is closely related to the mass
parameters in the correlator of the 4-currents from which the OPE
is calculated and the sum rules are derived.

A preliminary discussion of the mass dependence has been presented
in ref.~\cite{co},
where power corrections to the sum rule for a specific combination of the
helicities of pion Compton scattering have been presented.
An artificial mass dependence is generated
in the calculation of such corrections (this point is briefly illustrated in
section 4)
and we have argued, in the same work,
that this feature was not going to affect
the size of the analyticity region or the validity of the sum rule,
for small values of the mass.

Here we are going to give a general proof of these statements
together with an explicit description of the
analytical bound for the dispersion relation,
generated by a finite mass dependence
in the expansion of the correlator.

Such thresholds, absent in the case of form factors,
cannot be ignored in the case of $s$ and $t$ dependent dispersion
relations. This bound is given by a quadratic equation, and shows up
as a singularity of the lowest order spectral density (the vanishing
of its denominatror). We show in sections 3 and 4 that moderate values of
$s$ and $t$ keep the bound far away from the (finite) region in which
the OPE is calculated. A similar (mass-dependent)
bound appears for the diagrams describing the power corrections.
The presence of these bounds is justified by the fact that each
coefficient of the OPE has leading singularities (also called $Landau$
$surfaces$) generated when all the internal lines in the perturbative expansion
go on shell. For Compton scattering these surfaces corresponds to a
$u$ channel threshold. We show, however, that these singularities,
though important in a more general analysis (in mass dependent correlators),
disappear in the massless limit. Therefore, in this particular case,
a simplified structure of the OPE emerges,
which is fully exploited in the calculation of the radiative corrections.
We
are also able to write down a dispersive representation
of the main integrals (the coefficients of the OPE)
which is unbound in the plane of the two dispersive variables.
The proof is limited to the lowest order result.

In this simplified case we prove
that a limited applicability of Borel methods \cite{NR1,JS} (Appendix C)
is possible, and that the leading spectral density
 can be re-obtained elegantly
without using the usual Cutkosky rules.
This is consistent with our results of section 3.

The absence of these $u$-channel cuts (in the massless case)
in the lowest order correlator is crucial for the evaluation of the radiative
corrections. In fact we are looking for imaginary parts to the
spectral densities and, to next order
in perturbation theory, these cuts reappear in specific subdiagrams.
The analysis presented in sections 3 and 4 is therefore preliminary
to our subsequent discussion.
In section 5 we illustrate, using various results
of dispersion theory, how to organize the
calculation of the radiative corrections
to the spectral densities in a systematic way. The analysis of the
complex parts in the diagrammatic expansion of the spectral density,
is illustrated for all the independent set
of diagrams, and a simple $factorized$ structure (in terms of specific
$sub-cuts$) of the OPE emerges.

Results of dispersion theory to
lower order are then used to decide of the real or complex behaviour of such
functions. Remarkably, thanks to this simple factorized feature,
we show that such corrections can be evaluatedd
in closed form. For this purpose, we combine dimensional regularization
and light-cone methods, and show that the divergences in the OPE
can be isolated as poles in $\epsilon=n\,\,-\,\,4$.

Our conclusions are in section 6.
Appendices A-C cover technical derivations of sections 3-5, while
Appendix D illustrates, in a self contained way,
 how to extend Borel methods \cite{NR1} to
Compton scattering.

\section{Spectral densities for 4-point correlators}
Sum Rules relate the timelike region of a correlator to its spacelike part
by a dispersion relation \cite{NR1,JS,SVZ}.
%%%%%%%%%%%%%%%
In the case of pion Compton scattering, for instance, the
4-point correlator with non vanishing projection over the
invariant amplitudes (the helicities) of the process
is given by \cite{CRS}
\begin{eqnarray}
\Gamma_{\sigma\mu\nu\lambda}(p_1^2,p_2^2,s,t)&=&i\int{\rm d}^4x {\rm d}^4y
 {\rm d}^4z\exp
(-ip_1\cdot x+ip_2\cdot y-iq_1\cdot z)
\nonumber \\
& &\times \langle 0|T\left(\eta_{\sigma}(y)J_{\mu}(z)J_{\nu}(0)
\eta_{\lambda}^{\dagger}(x)\right)|0\rangle \; ,
\label{tp}
\end{eqnarray}
where
\begin{eqnarray}
J_{\mu}=\frac{2}{3}\bar{u}\gamma_{\mu}u -\frac{1}{3}\bar{d}\gamma_{\mu}d,
\;\;\;\;\;\;
\eta_{\sigma}=
\bar{u}\gamma_5\gamma_{\alpha}d
\label{jd}
\end{eqnarray}
are the electromagnetic and axial currents, respectively,
of up and down quarks.
 $q_1$ and $q_2$ are on shell moment carried by the two physically
polarized photons (see Fig.~1).

The two pion momenta are denoted by $p_1$ and $p_2$,
with $s_1=p_1^2$ and $s_2=p_2^2$ being the virtualities.

The invariant amplitudes of pion Compton scattering are then extracted
from the physical expansion of this correlator for timelike momenta
$p_1, p_2$.
The matrix element which interpolates with pion states is
given by \cite{CRS}
\beq
M_{\nu\lambda}= i\int d^4y   e^{-iq_1 y}
  \langle p_2|T\left(J_{\nu}(y)J_{\lambda}(0)\right) |p_1 \rangle\; ,
  \label{mnula}
\eeq
and can be related to the two helicities $H_1,H_2$
as
\beq
M^{\lambda\mu}= H_1(s,t) e^{(1)\lambda}e^{(1)\mu} +
H_2(s,t) e^{(2)\lambda}e^{(2)\mu},
\label{h1h2}
\eeq
where $e^{(1)}$ and $e^{(2)}$ are helicity vectors defined in \cite{CRS,co}.

The spectral densities appearing in the sum rules
are characterized by a resonance contribution at the pion pole
$\Delta^{res}(s_1,s_2,s,t)$,
 and by a continuum one $\Delta^{\rm pert}(s_1,s_2,s,t)$,
at large values of the virtualities $s_1$ and $s_2$ \cite{CRS,co}.
A dispersion relation than allows us to connect the spacelike region
of the 4-current
correlator in eq.~(\ref{jd}) to its timelike behaviour in
$p_1^2, p_2^2$ at fixed angle $-t/s$.

The resulting sum rule for $H_i(s,t)$ ($i=1,\,2$) can be written down
in terms of a local duality contribution and of power corrections of
quarks and gluons as

\beqa
&&{f_\pi}^2H_i(s,t)\left({s(s+t)\over -t}\right)=\nonumber \\
& &\hspace*{0.5cm}\left(\int_{0}^{s_0}ds_1\int_{0}^{s_0}ds_2
{\rho_i}^{\rm pert}+\frac{\alpha_s}{\pi}\langle G^2 \rangle
\int_{0}^{s_0}ds_1\int_{0}^{s_0}ds_2\rho_i^{\rm gluon}\right)
e^{-(s_1+s_2)/M^2} \nonumber \\
& &\hspace*{0.5cm}+C_i^{\rm quark}\pi\alpha_s
\langle (\bar{\psi}\psi)^2 \rangle\;.
\label{h2}
\eeqa

where $s_0$, the local duality interval, characterizes the resonant region
of the correlator.
The spectral densities $\rho^{\rm pert},\rho^{\rm quark},\rho^{\rm gluon}$
are calculated by the OPE. They are polynomial
in a variable $Q(s_1,s_2,t)$ which has a natural interpretation in the
light-cone frame of the pion' lines (see Appendix A).

The evaluation of all the lowest dimesnional contributions to the
OPE of the two helicities has been presented in ref.~\cite{H1H2}.

For instance, the covariant expression of $\rho_i^{\rm pert}$
is of the form
\beqa
&& \rho_i^{\rm pert}={5\over 24 \pi^2}{R_i(Q^2,s,s_1,s_2)\over
Q^4 (2 Q^2 -s)^2(4 Q^4 - s_1 s_2)^5 (2 Q^2 s - s_1 s_2)^2}\;,
\eeqa
where
\beqa
R_i(Q^2,s,s_1,s_2)=\sum_{n=0}^{15}a_{i,n}(s,s_1,s_2) Q^{2 n}.
\eeqa
The explicit forms of the coefficients $a_{i,n}$ can be found in
\cite{H1H2}.
The contributions from the lowest dimensional
vacuum condensates are also given in that work.

As shown below in eq.~(\ref{qsq}),
the expression of $Q(s_1,s_2,t)$ is non trivial
and although it can be viewed
as the basic momentum scale which characterizes the
power law fall off of the spectral densities (see \cite{co}),
(which are simple rational functions when expressed in this variable),
it is however not exactly on the same ground as  $s$ and $t$
in characterizing the "dimensional counting" of the sum rules
(as it is usually done in standard factorization \cite{BL}).
There are two main reasons for this:
1) $Q$ contains the sum of $both$ $s$ and $|t|$ at larger energies
(which in our case are supposed to be of the order of
5-7 $Gev^2$ or so \cite{coli})
2) it appears as a variable
(under the local duality interval)
and therefore it is not just a fixed external invariant.
The power law falloff of the spectral density, however,
shows that \cite{co,H1H2}, generically
\beq
\rho(s_1,s_2,s,t)\approx {C_1\over Q^6} + {C_2\over Q^8}, +...
\eeq
where $C_1$ , $C_2$, etc, are complicated functions of the pion virtualities
 and contain the input from the condensates.
Therefore, contributions from "higher-twists" are embodied
in an obvious form by the dispersion relation. There are some
specifc features of this approach which clearly differentiate it
from the perturbative picture: 1) to lowest order,
no gluons are involved,  2) the power law falloff is much more
suppressed (a coherence effect)
compared to the perturbative one and 3) the sum rule
prediction starts
dominating compared to the perturbative one toward the region of forward
scattering. Features 1) and 2) are known from the literature on
sum rules for form factors. However, in this latter case,
the contributions from the condensates grows as a power of "Q" and soon
overtakes the (decaying) contribution from the perturbative coefficient of the
OPE. In our case, instead, a suppression is generated -for scattering at
fixed angle- at the photon vertex by a virtual quark line connecting the two
photons and eliminates this growth. In fact , compared to form factors,
a dependence on both $s$ and $t$ is now availiable in the sum rule.
The shortcome of this, however, is that the stability analysis
of the sum rules (in both $s$ and $t$) becomes much more complex \cite{H1H2}.

In order to describe, for instance, interference effects in
 photoproduction processes ($\gamma \,\,\gamma \to\pi^+ \pi^-$)
and in other related processes by a sum rule to order $\alpha_s$,
one needs the imaginary parts of the radiative corrections to the
spectral densities. The real parts are also, as a matter of fact, needed,
however an actual calculation shows that these are suppressed by
a factor $1/Q^4$ compared to the imaginary ones.
The argument is quite simple and can be better understood from
a reading of section 5. Basically, imaginary parts are generated
by pinched diagrams containing 5-particle cuts, while real parts require only
 4-particle cuts. This implies that more propagators of large virtuality
($1/Q^2$) appear in the integrals describing real discontinuities,
compared to those contributing to the imaginary ones.
One can also define suitable observable were the contributions from the
imaginary parts are enhanced compared to the real ones.
%Just to give an example, let's start considering
%the differential cross section \cite{H1H2} for pion photoproduction,
%given by \cite{H1H2}
%\beq
%{d\sigma\over d\,\,cos\,\,\theta}{|{\cal M}|^2\over 32 \pi s},
%\eeq
%where {\cal M} is the invariant amplitude fo a specific reaction
%and $cos\,\,\theta=(t\,\,-\,\,u)/s$ is the center of mass scattering angle.
%For instance, in the case of the process $\gamma \,\,\gamma \to\pi\pi$

%\beq
%|{\cal M_1}|^2={|H_1|^2 + |H_2|^2\over 2},
%\eeq
%while for $\gamma_R\,\,\gamma_R\to \pi^+\,\,\pi^-$

%\beq
%|{\cal M_2}|^2={|H_1 -H_2|^2\over 2}.
%\eeq
%Then the difference
Here is a simple example of how this may be happen.

If we include radiative corrections, $H_1$ and $H_2$
are determined by the $complex$ spectral densities

\beq
\rho_i(s_1,s_2,s,t)=\rho_i^{(0)} + \alpha_s \left(\rho_i^{(1)} + i\,\,
\rho_i^{(2)}\right) + \rho_i^{pc}
\eeq
where the suffix $i$ runs from 1 to 2. $\rho^{pc}$ is a short
notation for the contributions coming from the power corrections.
It is not difficult to choose specific observable which
are more sensitive to these complex parts than to the real ones.
For example, simple counting
arguments are sufficient to show that the difference

\beq
\sigma\equiv\sigma(\gamma \,\,\gamma \to\pi\pi)\,-
\sigma(\gamma_R\,\,\gamma_R\to \pi^+\,\,\pi^-)
\eeq

is dominated by the lowest order spectral density and by the purely imaginary
parts of their  radiative corrections $(\rho_{\sigma}\approx
\rho_1^{(0)}\rho_2^{(0)} + \rho_1^{(2)} \rho_2^{(2)})$.
As we are going to show in the next few sections, the imaginary parts
of the sum rules of meson photoproduction can be classified according to
a well defined factorized structure and are calculable in a close form.

Notice that the sum rule studied so far have been calculated in the
approximation of
massless quarks in the expansion of the correlator.
Although this is clearly apppropriate in the case of the pion,
may not be appropriate for the investigation of heavier mesons.

In fact, the method can be easily applied to all the other
reactions obtained by simple crossing (in the
$s$, $t$ and $u$ channels) of the Compton scattering amplitude, and
in particular,
to processes of the form $\gamma+ \gamma \to A+ \bar{A}$,
where $A$ can be, in principle, any meson.

The evaluation of the sum rule, for these more general reactions,
encounters a special difficulty related to the presence of
a mass $m$ in the diagrammatic expansion.
As we are going to point out in the next few sections, this
problem does not show up in the case of the form factor
since the spectral density - at least to lowest order -
is mass independent \cite{JS}.
Such issues, although of technical nature,
deserve special attention in the sum rule context.
Our discussion, here, is ground-breaking
since there are no available methods in the calculation of the OPE
of the double spectral densities beyond 2- and 3-point functions.
 In particular we point out that
the use of the Borel transform \cite{NR1}
in the isolation of such spectral functions must be handled with a certain
care when the rank of the correlator increases (such as for 2 photon
processes).
We refer to the discussion in Appendix C
for a presentation of the Borel method and its partial generalization to
the evaluation of the lowest order spectral functions for Compton scattering.

In the case of the form factor, the radiative corrections, even in the
massive case, can be obtained by the Borel method \cite{braun}
which is, in fact, the easiest way to
evaluate such contributions.
As we are going to discuss next, in Compton scattering,
these methods, \cite{NR1}
which rely on simple properties of the Borel transform
and on the existence of unbound dispersion relations
for the corresponding Feynman diagrams, are of limited use.
Although the particular kinematics chosen in
ref. \cite{CRS} for the analysis of this reaction by sum rules
(the requirement of treating on-shell physical photons)
brings our treatment closer
to the one in the form factor case, yet new analytical features emerge in the
leading order spectral densities of these processes
and in their power corrections \cite{co}.
Therefore, we proceed in sections 4 and 5 to develope our own approach
to the identification of the complex contributions to the sum rules.

Our discussion, in the next sections, relies only on
the analysis of massless and massive $scalar$
amplitudes. There is no need, in fact, to consider the full
expression of the relevant
diagrams for processes of these type
(and this is true also in the form factor)
since the basic difficulties in the
analytic properties of the spectral densities
are all contained in the scalar amplitudes.
%%%%%%%%%%%%%%%%
\section{Massive correlators}

Here we start our discussion of the role of a mass dependence
by considering 3-point functions.
The general features of the sum rule method for the analysis of
form factors can be found in refs.~\cite{NR1} and \cite{JS}.
In this case \cite{JS}, the region of analyticity
(for any 3-current correlator
which interpolates with a specific form factor)
is unbound and the dispersion integral extends up to
infinity. In fact the basic spectral density is obtained
from the integral (the $triangle$ singularity) \cite{Eden,JS}
\beq
\Delta_3 (s_1,s_2,t)=
\int d^4 k\delta_+(k^2-m_1^2)\delta_+((p_1-k)^2-m_2^2)\delta_+((p_1-k +
q_1)^2-m_3^2),
\label {trian}
\eeq
 The propagators
of the cut lines have been replaced, as usual, by "positive" delta functions
\beq
{1\over k^2 -m^2}\to -2 \pi i \delta_+(k^2-m^2),
\eeq
as prescribed by Cutkosky rules.
The relevant diagram is depicted in Fig.~1b.
It gives the leading spectral function, for form factors, to lowest order
in $\alpha_s$ \cite{JS}. The dashed lines describe the usual cutting rules for
the propagators. The spectral density
 for such a diagram takes the form (with $p_1^2=s_1,\, p_2^2=s_2$)
\beqa
\label{delta3}
\Delta_3(s_1,s_2,t)=
{\pi \theta(s_1-4 m^2)\theta(s_2-4 m^2)
\over 2 \left(\lambda(s_1,s_2,t)\right)^{1/2}}
\eeqa
where
\beq
\lambda(s_1,s_2,t)= \left( (s_1 + s_2 -t)^2 - 4 s_1 s_2)\right)^{1/2}.
\label{mand}
\eeq
is the Mandelstam function.
The evaluation of (\ref{trian}) can be easily carried out in the Breit frame of
the two (pion) lines
characterized by the virtualities $p_1^2$ and $p_2^2$.
It is also easy to realize that (\ref{delta3}) does not develope
any additional singularity
for positive  $s_1,\ s_2$, and negative $t$ ($t=(p_2 - p_1)^2$ fixed).

Another important observation, in the case of form factors,
is the independence of the discontinuity eq.~(\ref{trian})
from any mass in the propagator (see also \cite{Eden}).
Eq.~(\ref{trian}) is in fact regular
for any $s_1$ and $s_2$ over the entire complex plane of each of these
variables.
This allows us to write down a double dispersive representation of
the massless triangle diagram in a rather straightforward way
as simply as
\beqa
T_3(p_1^2,p_2^2,t)&=&\int {d^4 k \over k^2 (p_1-k)^2 (p_2-k)^2}\nonumber \\
&&=\int_{0}^{\infty}ds_1\int_{0}^{\infty}ds_2{ \Delta_3(s_1,s_2,t)\over
(s_1-p_1^2)(s_2-p_2^2)} +\,\, subtr. \nonumber \\
\label{key}
\eeqa
In eq.~(\ref{key}) we have omitted subtractions
and single dispersive contributions to
this integral. These two features  1) mass independence of the double spectral
function  and 2) its regularity over the entire $s_1$ and $s_2$ complex planes,
are not preserved in Compton scattering, except for the massless case.
Additional care is therefore required when we switch on a mass dependence in
the
basic diagrams, since new thresholds related to the $u$-cut in the dispersion
variables automatically appears in the complex planes of the two virtualities
$s_1$ and $s_2$.

To be specific, let's consider the full contribution to the box diagram
with a momentum flow chosen as in Fig.~1b, and for simplicity,
let's restrict our considerations to the scalar case.
We consider the following 4-point function
\beq
T_4=\int {d^4 k \over k^2 (p_1-k)^2 (p_1-k+q_1)^2 (p_2-k)^2}
\label{t4}
\eeq
with massless propagators.
At fixed angle and with
\beq
s+t+u=p_1^2 + p_2^2 \,\,\,\,\, s>0,\,\,t<0, \,\,u<0
\label{con}
\eeq
and moderately large $s,t,u$, assuming the existence of a region of
analyticity of moderate size in $s_1$ and $s_2$ (see Fig.~2),
 we can write down a spectral representation for such integral
of the form
\beq
\label{t4g}
T_4=-{1\over 4 \pi^2}\int_{\gamma 1} ds_1\int_{\gamma 2}ds_2
{\Delta (s_1,s_2,s,t)\over (s_1 - p_1^2)(s_2 - p_2^2)}
\eeq
where the contours $\gamma_{1,2}$, both of a radius $\lambda^2$,
are again chosen as in Fig.~2.
If we introduce a double spectral function $\Delta(s_1,s_2,s,t)$,
we can rewrite $T_4$ as
\beq
T_4(p_1^2,p_2^2,s,t)=-{1\over 4
\pi^2}\int_{0}^{\lambda^2}ds_1\int_{0}^{\lambda^2}ds_2
{\Delta(s_1,s_2,s,t)\over (s_1-p_1^2)(s_2-p_2)^2} + . . .
\label{t4la}
\eeq
where the neglected pieces involve a complex part of the contour
(the $background$ contribution).

It has been shown in ref. \cite{CRS} that
 the leading perturbative spectral function
can be obtained, for this diagram, with the conditions  on $s$ and $t$
given by (\ref{con}), by the 3-cut integral (see Fig.~1b)
\beq
J(p_1^2,p_2^2,s,t,m=0)=\int
d^4k{\delta_+(k^2)\delta_+((p_1-k)^2)\delta_+((p_2-k)^2)\over
(p_1-k+q_1)^2}.
\label{Jzero}
\eeq
The evaluation of (\ref{Jzero}) has also been discussed in \cite{CRS} and the
answer,
for the associated spectral density, turns out to be rather
simple
 \beqa
\label{set}
\Delta(s_1,s_2,s,t)&=&(-2\pi i)^3 J(p_1^2,p_2^2,s,t) \nonumber \\
                               &=& -{4 i\pi^4\over s t}.
\eeqa
Compared to the form factor case (see eq.~(\ref{delta3})),
the discontinuity
along the cut, $\Delta(s_1,s_2,s,t)$, as given by (\ref{set}) is
$s_1$ and $s_2$ independent.
This results seems to be oversimplified.
However such a simplification is due to the fact that
1) we have neglected the quark masses in eq.~(\ref{t4})
and 2) we have not considered other subleading cuts
in the  in the evaluation of (\ref{set}) which are suppressed by power of
$s, t $ or $u$, compared to the leading result (the single dispersive
contribution).

As we are going to show next, even in the simpler conditions of
a kinematics of Compton type, a dispersion relation in $s_1$ and $s_2$
has unobvious features, not evident from eq.~(\ref{set}).

In fact, while for a vertex function there are only 3 kind of cuts and 3
independent
variables (say $s_1$, $s_2$ and $t$), one can vary any of them and  keep the
others below their thresholds, obtaining a dispersion integral
extended up to infinity, in the case of the box diagram we have
7 cuts (respectively in $s,t,u,$ plus the 4 external virtualities)
and only 6 independent variables. The maximum number of variables
which can be kept fixed is five.

A variation in the 6-th variable will then
automatically affect the 7-th one. In our case, in particular, only 5 variables
play a role since 2 external lines are fixed to be massless (photons of
$q_1^2=q_2^2=0$)
from the very beginning.

Let's suppose that we fix $s$, $t$ and $s_2$ and vary $s_1$.
Then the 5-th variable $u=s_1 +s_2 - s -t$ can cross the threshold ($u=0$)
(see Fig.~2), for a sufficiently large $s_1$, and become positive.

The diagram corresponding to this threshold is shown in Fig.~3.
For $m=0$ and at fixed angle, however,
such discontinuity doesn't play any role, as far as
$t$ is negative.

The proof of this last statement goes as follows.
In Fig.~3 (for vanishing $m$)
 the only possibility of having lines 1, 2 and 3 on shell at the same
time, giving a non zero contribution to this diagram,
 is when these lines are collinear. For the same reason, lines 3, 4 and 5
must be collinear too. Therefore 1 and 5 must be collinear
(i.e. $q_1\cdot q_2=0$), which implies that we
 must restrict ourselves to consider forward
scattering from the very beginning.
The requirement of working at fixed (and non forward or
backward)
angle ($-t/s$ fixed)
allows us to exclude such configurations.

A rapid look at the leading spectral density given by
eq.~(\ref{set}) shows that this is indeed consistent with
the expression that we obtained before for $\Delta(s_1,s_2,s,t)$,
 since at $t=0$ our
spectral density becomes singular.

Therefore, as far as $t$ remains in the physical region (and $m=0$),
 the contour representation of the
box diagram eq.~(\ref{t4g}) (Fig.~2)
can be sent to infinity and the Cauchy integral replaced
by its double discontinuity
eq.~(\ref{set}), integrated over the positive semiaxis of $s_1$ and $s_2$

\beq
T_4 =-{1\over 4 \pi^2}\int_{0}^{\infty}\int_{0}^{\infty}ds_1 ds_2
{\Delta (s_1,s_2,s,t)\over (s_1-p_1^2)(s_2-p_2^2)} \,\,\,+\,\,\, subtr.
\label{ext}
\eeq
This unexpectedly simple result is a consequence of the kinematics chosen
in the derivation of the sum rule.
This result, as we are going to see,
 is no longer valid when an explicit mass dependence in the
diagrams is introduced.

Therefore, in the case $m=0$
and with on shell photons, it is also easy to write down the explicit
expression of the
dispersion integral in eq.~(\ref{ext}) as simply as

\beq
T(p_1^2,p_2^2,s,t,\mu)={1\over 2 s t (p_1^2- \mu^2)(p_2^2-\mu^2)}
Log\left({p_1^2-\mu^2\over p_2^2-\mu^2}\right) +\,\,\,\,\, subtr.,
\eeq
where we have chosen the same subtraction point $\mu$ for both
momenta $p_1$ and $p_2$ and neglected single dispersive parts in the integral.
It is important to observe
that the validity of a dispersion relation of the form given by eq.~(\ref{ext})
is crucial in order to proceed to the evaluation of the spectral density by
Borel techniques.

In order to illustrate how a bound on the
validity of the dispersion relation comes into play,
we have to reconsider the
lowest order spectral density in its more general form, this time
with a non zero mass in it.
Below we are going to prove 3 basic facts:

1) for a non zero mass $m$ the spectral density
 developes special singularities in the complex $s_1$ and $s_2$ planes;

2) the manifold spanned by the momenta which make the spectral density
singular is a Landau surface for a $u$-channel singularity;

3) in the limit of vanishing mass the surface is degenerate and the
spectral density remains regular, as shown above in eq.~(\ref{set}).

Let's proceed from point 1) and let's first discuss how a mass term is going to
bound the region of analyticity.

The leading spectral function - for a non vanishing $m$ -
is obtained, in this more general case, from the 3-particle cut integral
\beq
J(p_1^2,p_2^2,s,t,m)=\int d^4 k {\delta_+(k^2-m^2)\delta_+((p_1-k)^2-m^2)
\delta_+((p_2-k)^2 - m^2)\over (p_1-k+q_1)^2 - m^2},
\label{jem}
\eeq
and it can be expressed in the form
\[
\Delta(s_1,s_2,s,t,m) = (-2 \pi i)^3 J(s_1,s_2,s,t,m)
\]
\beq
=-{4 \pi^4 i \over\sqrt{-t (4 m^2 s_1 s_2 - 4 m^2 s s_2 -4 m^2 s s_1 + 4 m^2
s^2 + 4 m^2 s t - s^2 t)}},
\label{djem}
\eeq
which reproduces eq. (\ref{set}) in the $m=0$ case.

The evaluation of (\ref{djem}) is not obvious and can be performed following
the steps discussed in Appendix A, where we discuss a similar integral.
For two different masses, say $m_1$ and $m_2$, in the box diagram,
an expression of $\Delta(s_1,s_2,s,t,m_1,m_2)$ in terms of
$s$ and $t$, as nice as eq.~(\ref{djem}),
is very difficult to obtain. In general, rational powers of
the  Mandelstam function
$\lambda(s_1,s_2,t)$ will appear and cannot be eliminated by a
simple reshuffling of the indipendendent variables $s_1,s_2,s,$ and $t$.

Differently from eq.~(\ref{set}) which is mass independent and globally a
regular function of the two pion' virtualities,
eq. (\ref{djem}) shows, from its denominator,
 that the spectral function developes a singularity at
$s_2$ dependent positions in the $s_1$ plane (and $viceversa$),
as it is expected in general.

We intend to elaborate some more on the origin of this singularity which
is of $u$-type, since it is generated
when all the 4 internal lines in the massive box diagram go on shell
(see the discussion in \cite{Remiddi}).

We remind here that the conditions $q_1^2,q_2^2>0$
has to be satisfied, in general,
 in order to get contribution from such
discontinuity (for the box diagram),
since a massless photon cannot decay into two massless
particles at $t<0$. The fact that such singularity reappears,
also in the approximation of prompt photon emission $(q_1^2=q_2^2=0)$
imposed from the very first stage of our calculations, shows that
dispersion relations in two variables, for 4-point functions,
are very difficult to handle on a very general base even for the
simple box diagram.
%\footnote{I thank Alan White for clarifying this point to me}

Neverthless we can easily prove that
$\Delta(s_1,s_2,s,t,m)$, as given by eq.~(\ref{djem}),
 becomes singular
when the invariants $s_1$ and $s_2$ satisfy the Landau equation for
obtaining a singularity of $u$-type.

The proof goes as follows.
The leading singularity of the box diagram can be described geometrically by
 a simple dual diagram (generated by the momenta of the original box diagram)
\cite{Eden}
whose volume is imposed to be zero.
This condition can be usually expressed in a more simple form in terms of
some canonical variables $y_{i,j}$, proportional to the
 scalar product of the internal momenta in the diagram. In our case
(for equal internal masses $m$) such variables are

\beqa
&& y_{12}=-1\,\,\,\,\,\, \,\,\,\,y_{13}={t - 4 m^2\over 2 m^2} \nonumber \\
&& y_{23}=-1\,\,\,\,\,\,\,\,\,\,\, y_{14}={s_1 - 2 m^2\over 2 m^2}\nonumber \\
&& y_{24}={s- 2 m^2\over 2 m^2}\,\,\,\,\,\,\,\, y_{34}={s_2-2 m^2\over 2 m^2}.
\label{ys}
\eeqa

The equation of the Landau surface corresponding to Fig.~3 is in general given
by a polynomial in $y_{i j}$ \cite{Eden}

\beqa
&& 1 - {{{ y_{12}}}^2} - {{{ y_{13}}}^2} - {{{ y_{14}}}^2} -
  2 { y_{12}} y_{13} { y_{23}} \nonumber \\
&& - {{{ y_{23}}}^2} +
  {{{ y_{14}}}^2} {{{ y_{23}}}^2} - 2 { y_{12}} { y_{14}} { y_{24}} -
  2 { y_{13}} { y_{14}} { y_{23}} { y_{24}} - {{{ y_{24}}}^2} +
  {{{ y_{13}}}^2} {{{ y_{24}}}^2} \nonumber \\
&& - 2 { y_{13}} { y_{14}} { y_{34}} -
  2 { y_{12}} { y_{14}} { y_{23}} {y_{34}} -
  2 { y_{12}} { y13} { y_{24}} { y_{34}} -
  2 { y_{23}} { y_{24}} { y_{34}} - {{{ y_{34}}}^2} +
  {{{ y_{12}}}^2} {{{ y_{34}}}^2}=0
\label{ys1}
\eeqa

Inserting (\ref{ys}) into (\ref{ys1}) one gets
the expression
\beq
-{t\over m^4} (4 m^2 s_1 s_2 - 4 m^2 s s_2 -4 m^2 s s_1 + 4 m^2
s^2 + 4 m^2 s t - s^2 t)=0
\label{sing}
\eeq
which clearly coincides with the singularity of the massive spectral density
in eq.~(\ref{djem}).

It is important to realize, however,
that such singularity
is far enough from the region of analyticity in which the dispersion relations
are enforced.

In fact, the leading behaviour
of such densities, even for non vanishing $m$, is still given
(in eq.~\ref{djem})
by ${-1/(2 s t)}$, at large $s$ and $t$, and at fixed value of their
ratio $-t/s$.
It's easy to show from the same equation that such a threshold is located
at the typical values $s_1, s_2\approx s( 1 + O(m^2/s))$
of the two virtualities, reasonably far away from the
border of the contour of Fig.~2, for a dispersion relation in
$s_1$ and $s_2$ to be valid.

Whence we can summarize  our findings in this section as follows

1) We have explicitely described (by eq.~(\ref{sing})) the
location of the $u$ channel cut in the spectral density in the two complex
planes of $s_1$ and $s_2$ and 2) we have shown that the only parameter which
controls the position of this singularity is the mass in the correlation
function.

At large $s$ and $t$ the term $-s^2 \,\,t$ in eq.~(\ref{sing}) $dominates$
compared to all the other terms, even for values of the
virtualities large compared to the mass $m$ in the correlator,
but smaller compared to
$s$ and $t$. The limit $m\to 0$ can be studied from eq.~(\ref{sing})
by sending to zero the mass in the numerator of such equation independently
from the rest. The singularity surfaces degenerates, in this limit,
the spectral density becomes regular
and we reobtain eq.~(\ref{set}).
In the next section we are going to extend our reasoning to the diagrams of the
power corrections.

\section{Power Corrections}
In order to analize the analyticity properties of the diagrams which
appear in power corrections of gluonic type
(see Fig.~4) we first review our method of calculation of such terms
and show that in the massless case,
again, the $u=0$ threshold, present in these diagrams,
 has a location in the complex planes of
$s_1$ and $s_2$ which is parametrically controlled by the mass $m$
of the OPE and disappears in the massless limit.
We are then able to write down a dispersive representation of
the diagrams for the power corrections which can be extended
for $s_1$ and $s_2$ positive, ranging up to infinity.
The pattern is similar to what already discussed in the previous section.
Let's first observe that the power corrections can be obtained
by the insertion of a modified quark propagator in momentum space \cite{mallik}
of the form
\beqa
S(p) &=& {\not{p} +m\over p^2 - m^2}\,\, + \,\,{1\over 2} i
{(\gamma^{\alpha}\not{p}
\gamma^{\beta} G_{\alpha \beta}
 - m \gamma_{\alpha}G^{\alpha\beta}\gamma_{\beta})\over (p^2
- m^2)^2} \nonumber \\
&\ &  \quad \quad + {\pi^2 \langle G^2\rangle m\not{p} (m +\not{p})\over (p^2 -
m^2)^4},
\label{propagator}
\eeqa
which has quartic powers of momenta at the denominators.
$\langle G^2\rangle$ is the gluonic condensate.
Insertion of this propagator results in diagrams of the form
shown in Fig.~4, for which we need an explicit spectral representation

\beq
J(p_1^2,p_2^2,s,t)
\equiv \int {d^4 k \over (p_1-k)^2 \left( k^2\right)^2 \left( (p_2 -k)^2
\right)^2 (p_1 - k + q_1)^2} \nonumber
\eeq
\beq
\label{disc1}
=-{1\over 4 \pi^2}\int_{0}^{\lambda^2}ds_1 \int_{0}^{\lambda^2}ds_2\,{
\bar{\Delta}(s_1,s_2,s,t)\over (s_1 - p_1^2)(s_2-p_2^2)}.
\eeq
We are interested in retrieving the explicit expression of
$\bar{\Delta}(s_1,s_2,s,t)$ for these diagrams.

For this purpose, as discussed in \cite{co},
it is convenient to introduce two auxiliary masses $m_1$ and $m_2$
and consider the more general spectral representation of the following
amplitude

\beqa
&& J(p_1^2,p_2^2,s,t,m_1,m_2) \nonumber \\
&& = \int {d^4k  \over
(p_1-k)^2 (k^2-m_1^2)^2
\left((p_2 -k)^2 - m_2^2\right)^2 (p_1 - k + q_1)^2} \nonumber \\
&& ={\partial\over \partial m_1^2}{\partial\over \partial m_2^2}
\tilde{J}(p_1^2,p_2^2,s,t,m_1,m_2),
\label{factorm}
\eeqa

where we have defined (see Fig.~4 a)
\beq
\tilde{J}(p_1^2,p_2^2,s,t,m_1,m_2)=\int
{d^4k\over (k^2 - m_1^2)\left( (p_2-k)^2 - m_2^2\right) (p_1-k)^2 (p_1-k +
q_1)^2}.\
\label{jmo}
\eeq
and factorized two derivatives respect to the masses $m_1$ and $m_2$
in eq.~(\ref{factorm}).

The basic idea of the method used in \cite{co} for the calculation of
such corrections is to first approximate eq.~(\ref{jmo}) by its double
dispersive part
\beq
\label{jm1}
\tilde{J}(p_1^2,p_2^2,s,t,m_1,m_2)=-{1\over 4
\pi^2}\int_{0}^{\lambda^2}ds_1\int_{0}^{\lambda^2}ds_2
\,{\tilde{\Delta}(s_1,s_2,s,t,m_1,m_2)\over (s_1 - p_1^2)(s_2 - p_2^2)}\
\eeq
with a spectral density given by
\beqa
&&\Delta^{scalar}(s_1,s_2,s,t,m_1,m_2) \nonumber \\
&& =(-2\pi i)^3\int d^4k\,
{\delta_+(k^2-m_1^2)\delta_+((p_2-k)^2 - m_2^2)
\delta_+((p_1-k)^2)\over (p_1-k +q_1^2)}.
\label{disco}
\eeqa
$\lambda^2$ is the radius of a finite contour in each plane of $s_1$
and $s_2$.
Then, the dispersive part of the scalar integral associated to
eq.~(\ref{disc1})
is obtained by setting $m_1$ and $m_2$ to zero
according to the relation

\beq
\label{disc2}
\bar{\Delta}(s_1,s_2,s,t)\equiv \left({\partial\over \partial
m_1^2}{\partial\over
 \partial m_2^2}\Delta^{scalar}(s_1,s_2,s,t,m_1,m_2)\right)_{m_1,m_2=0}.
\eeq

Evaluation of (\ref{disco}) gives
\beq
\label{delm}
\tilde{\Delta}(s_1,s_2,s,t,m_1,m_2)=(-2\pi i)^3 {\pi\over 2 \delta W(m_1,m_2)}
\eeq
with
\beqa
W(m_1,m_2) &=& - 4 m_1^2 (Q^2)^2 + 2 m_2^2 Q^2 s + 4 (Q^2)^2 s + 2 m_1^2 Q^2
s_1 -2 m_2^2 Q^2
s_1\nonumber \\
& & -2 Q^2 s s_1 + 2 m_1^2 Q^2 s_2 - 2 Q^2 s s_2 - m_1^2 s_1 s_2 + s s_1 s_2
\label{w}
\eeqa

 $Q^2$ is given by
\beq
\label{qquadro}
Q^2={1\over 4} (s_1+s_2-t + \lambda^{1/2}(s_1,s_2,t)) =
{1 \over 4} (s+u+\lambda^{1/2}(s_1,s_2,t))\ ,
\label{qsq}
\eeq
with

\beqa
\lambda(s_1,s_2,t)&=& s_1^2 + s_2^2 + t^2
-2 s_1 t -2 s_2 t - 2 s_1 s_2\nonumber \\
&=& {\left(4 Q^4 - s_1 s_2\over 2 Q^2\right)^2}.
\label{deltadef}
\eeqa
%$\lambda(s_1,s_2,t)$ is the usual Mandelstam funtion.
As described in Apppendix A, $Q$ is the large light-cone component in the
"plus" direction of the incoming momentum $p_1$.
Specifically,
we expand all the momenta in the Sudakov base
\beq
p_1=Q {n^+} +{s_1\over 2 Q} n^-,\,\,\,\,             p_2={s_2\over 2
Q} n^+ + Q n^-,
\eeq
where $n^+$ and $n^-$ are light-cone momenta, such that
${n^+}^2 ={n^-}^2=0,     n^+\cdot n^-=0. $
In our conventions
\beq
n^+= {1\over \sqrt{2}}(1,{\bf 0}_\perp,1)), \,\,\,\,
n^-={1\over \sqrt{2} }(1,{\bf 0}_\perp,-1).
\eeq

The momentum transfer  $t$ can be expressed
in terms of $Q^2$, which can be viewed
approximately
as a large "parameter" in the scattering
process, by the relation

\beq
t={{-\left( \left( 2 {Q^2} - { s_1} \right)
       \left( 2 {Q^2} - { s_2} \right)  \right) }\over {2 {Q^2}}}
\label{ti}
\eeq

Notice that $t=-2 Q^2$ for $s_1=s_2=0$.

Now, simple manipulation of eqs.~(\ref{delm}) and eq.~(\ref{disc2})
give
 \beqa
\bar{\Delta}(s_1,s_2,s,t)={-32 \pi^4i(Q^2)^2 (s_1 -s)(-4 (Q^2)^2 +2 Q^2 s_1 - s
s_1
+ 2 Q^2 s_2)\over s^3 (2 Q^2 - s_1)^3 (-2 Q^2 + s_2)^3}\ .
\label{disc3}
\eeqa

We now want to demonstrate that

1) the only singularities which bound
the dispersion integrals (\ref{jmo}) for the power corrections are of $u$-type

2) they disappear in the massless limit.

For this purpose, let's build on the results of the previous section by showing
that,
even for diagrams with gluonic insertions (Fig.~4)
the singularity surface of $u$-type
coincides with the singularity surface of the spectral density
eq.~(\ref{delm}).
In particular,
 all the considerations made in the previous section
 for the case of the leading order
spectral density - regarding the size of the region of analyticity of the
correlator
and the position of the $u$-cut - remain true also in this
more general case. It is also important to observe that this is consistent
with the structure of eq.~(\ref{factorm}) since the two derivatives which have
been factorized in that equation (for power corrections we deal with
$quartic$ propagators)
do not change the location and the nature
of the singularities of the integral given by eq.~(\ref{jmo}).

The discussion that follows
 also confirms the validity of the method developed in ref.~\cite{co}
for the evaluation of the power corrections in Compton processes.

Specifically we are going to show that

1) the singularity of $u$ type in the power correction diagram (Fig.~4 b)
is equivalent to the vanishing of $W(m_1,m_2)$ which is given in eq.~(\ref{w}),
and appears in the denominator of the spectral
density eq.~(\ref{delm}) of the power corrections;

2) in the massless limit such singularity surface disappears and the
spectral density (\ref{delm}) turns into the regular function (\ref{disc3}).

Similarly to what already discussed in the previous section,
let's define the parameters $y_{ij}$ for this new diagram, as we have done
in the previous section,
 which are helpful to describe such a singularity surface

\beqa
&& y_{12}=-{(m_3^2 +m_4^2)\over 2 m_3 m_4}
\,\,\,\,\,\, \,\,\,\,y_{13}={t - m_3^2-m_2^2\over 2 m_3 m_2}, \nonumber \\
&& y_{23}=-{m_4^2 + m_2^2\over 2 m_4 m_2},
\,\,\,\,\,\,\,\,\,\,\, y_{14}={s_1 -  m_3^2-m_1^2\over 2 m_1 m_3},\nonumber \\
&& y_{24}={s- m_4^2-m_2^2\over 2 m_2 m_4},\,\,\,\,\,\,\,\,
y_{34}={s_2-m_2^2-m_1^2\over 2 m_1 m_2}.
\label{ys2}
\eeqa
The equation defining the  Landau surface for diagrams of
 box typ, even in the presence of quartic propagators,
 is still given by eq.~(\ref{ys1}).
It is possible to rearrange
the equation of such a surface in the form

\beq
{P(Q^2,m_1,m_2,m_3,m_4)\over 64 m_1^2 m_2^2 m_3^2 m_4^2 Q^2}=0
\label{Pm}
\eeq
where $m_i$ is the mass of the $i-th$ internal line, and $P(Q^2,m_i)$ is
a complicated polynomial in the masses $m_i$ and in the momentum $Q$,
whose expression is too lengthy to be given here.
Although the calculations are quite involved,
neverthless it is possible to show that in
the limit $m_3,m_4\to 0$ eq.~(\ref{Pm}) simplifies since

\beq
P(Q^2,m_1,m_2,m_3,m_4)\to W(m_1,m_2)^2
\eeq
and becomes
\beq
    W(m_1,m_2)^2=0,
\eeq
with $W(m_1,m_2)$ given by eq.~(\ref{w})
and clearly coincides with the singularity surface of eq.~(\ref{delm}).

Therefore we have shown that in the massive case - even for different
masses - and, in particular, in the diagrams of the gluonic
power correction, the spectral density developes singularities of
$u$-type as conjectured in ref.~\cite{CRS}.
Such singularities are absent for vanishing $m_i$ and the spectral
integral, in this specific case,
 can be extended to the entire positive axis of $s_1$ and $s_2$.
Therefore, we can extend the results of ref.~\cite{co} and write down
a new dispersive representation of the power correction integral
eq.~(\ref{disc1}) for $s_1,s_2>0$ as

\beq
J(p_1^2,p_2^2,s,t)
=-{1\over 4 \pi^2}\int_{0}^{\infty}ds_1 \int_{0}^{\infty}ds_2\,{
\bar{\Delta}(s_1,s_2,s,t)\over (s_1 - p_1^2)(s_2-p_2^2)}.
\eeq
where we have omitted, again, single dispersive contributions.
A check on the consistency of our results can be obtained immediately
since $\bar{\Delta}(s_1,s_2,s,t)$ can be rewritten
in the form
\beq
\bar{\Delta}(s_1,s_2,s,t)={32 \pi^4 i (s_1-s)(s s_1 -2 s_1 s_2 - s_1 t
- s_2 t + t^2 -2 t \lambda^{1/2}(s_1,s_2,s,t)\over
(s_1 + s_2 -t + \lambda(s_1,s_2,s,t)) s^3 t^3},
\label{delclean}
\eeq
and the only singularity displayed by (\ref{delclean}) is a pole
at $t=0$. For scattering at fixed angle this pole doesn't play any role.

\section{Radiative Corrections}
In this section we develop methods for the calculation of
the radiative corrections to the lowest order sum rules.
Although we are not going to discuss the complete evaluation of
these terms to the actual case of pion photoproduction,
which we leave as a future investigation, here we will describe methods
which may have application to
a large class of reactions of Compton type and are easily generalizable
to the proton case. The method used is simply based
on the observation that
the complex contributions to the spectral densities can be "decomposed"
in specific subdiagrams which involve 2- and 3-particle cuts $times$
the lowest order result.
After this observation, we are then able to treat these contributions in a
systematic way.
Unfortunately, from a practical viewpoint,
 general schemes to treat such corrections
are not avaliable, since double dispersion relations
involve complex spectral densities beyond leading order.
So, one has to resort on a case by case study.
In particular, the treatment of the infrared divergences in the
dispersion formalism, for simple dispersive integrals,
 and the problems which naturally come from
its application, are well known \cite{BMR}.
Infrared divergences are often regulated by introducing a fictitious mass
in the discontinuity function, as done, in some cases, for the form factor.
In our case such a problem is even more complicated to deal with, since
we are handling double dispersion integrals.
As we have learned from the previous sections, a mass term $bounds$
the spectral density and modifies its singularities. Particular care is
therefore required when trying to take the massless limit in infrared
sensitive integrals \cite{BMR}.

The method that we are going to illustrate
in this section has two main features, specifically

1) it combines the
good quality of the "Breit frame" (namely the fact that in the "Q"-variable
the spectral density is polynomial) with

2) dimensional regularization.

These two properties are crucial in order to do the calculations in close form.
Dimensional regularization, in particular, is used in a plane which is
transversal with respect to
the two (longitudinal) light-cone variables in the $n^{\pm}$
directions. The difference of the method,
compared to the lowest order calculation, appears only
in the angular variables, since the angular integral has now dimension $n-3$.
Given the fact that
the lowest order result spectral density is infrared safe,
the use of dimensional regularization as $n$ goes to 4, does not modify
the lowest order expression of the spectral density \cite{H1H2}
(see Appendix C).

General arguments
 regarding the infrared safety of
the sum rule method to Compton scattering have been presented in
ref.~\cite{CRS}.
In a direct approach, however, one has to be able to show explicitely that
the infrared divergences generated by the all possible cuts cancel.
We reserve the discussion of this important point elsewhere.
The treatment of the ultraviolet divergences, instead, is
much less difficult since can be controlled by simple subtractions.
In particular, for Borel sum rules, such difficulty is
largely absent due to the fast convergence of the spectral densities after
Borel transforms.
This discussion can be addressed in a better way in the
actual case of Compton scattering rather than in an auxiliary scalar theory,
as we are doing it now. In particular, the use of Ward identities
in a true gauge theory
simplifies drastically the treatment and can possibly
make such infrared cancellations more transparent.

 Notice that standard tools, such as the largest time equation,
commonly used in the calculation of single variable cuts
in diagrams with any number of external lines,
(see \cite{tho}) are of
very limited help in the classification of the double spectral densities
of Feynman integrals.

In order to make our discussion as clear as possible, we have decided to
treat one specific diagram with mass regulators and the remaining ones by
dimensional regularization. The treatment of this first diagram
(which is, however, $real$) is
quite involved, but it may serve illustrate quite well our general
approach, and how it is possible to
derive a set of basic rules according to which we can deduce
whether a particular cut
gives a $real$ or a complex contribution to the OPE.

For this purpose let's consider Fig.~5a
with the momentum (and energy) flow chosen as shown in this figure.

Let's first observe that the double imaginary part of
such diagram (Fig.~5b) can be split into the convolution of a
2-particle discontinuity in the $s'$ energy variable, specifically
$s'=(p_1-l)^2$ and in a 2-to-2 scattering amplitude ( the top of the diagram),
calculated at Born level, with on-shell external lines (Fig.~5c).

The 2-particle cut  (bottom part of Fig.~5c)
contributes to the total discontinuity through the sub-integral
\beqa
\Delta^{5b}(s_1,s_2,p_1\cdot l)&& \nonumber
\eeqa

\beqa
 \int d^4k {\delta_+(k^2-m^2)\delta_+((p_1-k+l)^2-\mu^2)\over
((p_1-k)^2-m^2)( (p_2-k)^2-m^2)}
&=&{s'^2\over 2 \pi D^{1/2}}Ln\left( {a\, c - b\, d \,\,cos\,\theta
+ D^{1/2}\over
a\, c - b\, d cos\,\theta - D^{1/2}}\right) \nonumber \\
\label{logtwo}
\eeqa

where
\beqa
&& a=s'^2-s' (s_1+\mu^2- m^2)-(\mu^2-m^2) s_1; \nonumber \\
&& b=\lambda^{1/2}(s',\mu^2,m^2)\,\,\lambda^{1/2}(s',s_1,0); \nonumber \\
&& c=s'^2-s' (s_2+\mu^2- m^2) + (\mu^2-m^2) s_2; \nonumber \\
&& d=\lambda^{1/2}(s',\mu^2,m^2)\,\,\lambda^{1/2}(0,s_2,s'); \nonumber \\
&& q_{12}=q_1-q_2; \nonumber \\
&& D=(a\, c - b \,d \,cos\,\theta)^2- (a^2-b^2)(c^2-d^2); \nonumber \\
&& cos\theta=s'/((s'-s_1)(s'-s_2)) (2 t + s'-s_1-s_2 -s_1 s_2/s'); \nonumber \\
&& s'=s_1 -2 p_1^+ l^- - 2 p_1^- l^+. \nonumber \\
\label{as}
\eeqa

% u'=s_2-t -2 p_1^+ l^- -2 {p_1}^- l^+;

and where $\mu$ and $m$ are regulator masses.
They allow us to keep under control the divergences of these sub-cuts.

In order to obtain  the full contribution of
 diagram  5b we have just to
convolute its top part, shown in Fig.~5c, with (\ref{logtwo}),
in the form
\beq
\int_{-\infty}^{\infty}dl^+\int_{-\infty}^{\infty}dl^-\int \,d^2 l_\perp
{\Delta_{5b}(s_1,s_2,l) \delta_+(l^2-m^2)\delta_+
((q_{12}+l)^2-m^2)\over (q_1+l)^2-m^2}.
\label{almostfinal}
\eeq

The integrals over the transversal components of $l$ and
on one of the two longitudinal
 components, say $l^-$, can be done elementarily
(since $q_{12}\cdot l_\perp =0$)
 while the $l^+$ integrals is cut from above and from below by the
conditions on the momentum flow. The analysis of these conditions
is algebraically very involved and is briefly sketched in Appendix B.

It is possible to simplify eq.~(\ref{almostfinal}) even further.

For this purpose it is crucial to observe that, in the Breit frame,
eq.~(\ref{logtwo}) has no dependence on the angular variables of the
transversal plane, which have been
introduced by the Sudakov decomposition of the internal momentum
$l$
\beq
l=l^+ n^+ + l^- n^- + l_\perp .
\eeq

After having performed the integration over
the transversal variables, the final result for the diagram of
Fig.~5b can be expressed in the form

\beqa
I_{5b} &=&{\pi\over 4 q_{12}^+}\int_{\mu_1}^{\mu_4}dl^+ {\Delta_2(l^+,l_m)\over
\left( (q_1^+ l_m + q_1^- l^+)^2 -
2 q_1^+ q_1^-(2 l^+ l_m - m^2) \right)^{1/2}}
\label{derivation}
\eeqa
where
\beq
l_m\equiv -{(q_{12}^2 +2 q_{12}^- l^+)\over 2 q_{12}^+}.
\label{lm}
\eeq
The explicit expressions of $\mu_1$ and $\mu_4$,
are given in Appendix B, together with a brief discussion.

The contribution from diagram 5b is, however, real - for very small values of
the regulator masses $m$ and $\mu$ -  and therefore cannot be responsible
for any interference.
In order to illustrate this last point, let's show first that
the integrand in (\ref{derivation})
is a regular function of $l^+$ in the interval
of integration. Our analysis is exact in the limit of $m\to 0$.
An independent check of our reasoning will also be discussed below.
Let's rewrite (\ref{lm}) as
\beq
l_m= {A_1 l^+ + B_1\over (2 Q (2 Q^2 - s_2)(4 Q^4 - s_1 s_2)^2},
\eeq
where
\beqa
A_1 &=& 2 Q (2 Q^2 - s_1)(4 Q^4 - s_1 s_2)^2,\nonumber \\
B_1 &=& (-2 Q^2 + s_1)(-2 Q^2 + s_2)
( -16 Q^8 + 16 Q^6 s -8 Q^4 s^2 +4 Q^2 s s_1 s_2 - s_1^2 s_2^2). \nonumber \\
\eeqa
Inserting this result in the denominator of eq.~(\ref{derivation})
we obtain
\beq
{(q_1^+ l_m + q_1^- l^+)}^2 -2 q_1^+ q_1^- (2 l^+ l_m - m^2)=
 {N_1 {l^+}^2 + N_2 l^+ + N_3\over 4 Q^2 (4 Q^4 - s_1 s_2)^6},
\label{quadra1}
\eeq
where
\beqa
&& N_1=(2 Q^2 - s_1)(4 Q^4 - s_1 s_2)^6 \nonumber \\
&& N_2=2\, Q\, (s - 2 Q^2)(2 Q^2 - s_1)^2(2 Q^2 - s_2)(4 Q^4 - s_1 s_2)^3
\nonumber \\
&&\,\,\,\,\,\,\, \times (16 Q^8 -16 Q^6 s +8 Q^4 s^2 -4 Q^2 s s_1 s_2 + s_1^2
s_2^2)
\nonumber \\
&& N_3=m^2 4 Q^2(s-2 Q^2)(-2 q^2 + s_1)(-2 Q^2 +s_2)
(4 Q^4-s_1 s_2)^4(2 Q^2 s -s_1 s_2).
\label{quadra2}
\eeqa
 In the limit $m\to 0$ the quadratic form (\ref{quadra1}) has only complex
roots, since its discriminant vanishes.
Therefore there are no singularities of the integrand of (\ref{derivation})
in the variable $l^+$.
A small finite mass term does not modify our conclusions, since it is possible
to show, even in this case, that the two (now real) roots still
lay outside the integration region in $l^+$.

An alternative, more direct proof
of this fact can be obtained in the following way.
It is easy to realize that
any (eventual) additional imaginary part in (\ref{derivation})
is generated when the residual propagator of momentum $p_1\,\,+\,\,l$
(see also eq.~(\ref{almostfinal})) goes on-shell.
The discussion outlined in section 3 shows that this is indeed not possible
in the limit of a small mass m, unless $t=0$. Therefore the integrand is
regular.

After these considerations,
we are allowed to do a simple counting of the complex
factors of $i$, for this specific diagram,
as a check of our conclusions: we insert of a factor  $i^4$
(from the 4 single-particle cuts)
times a factor of $i^2$ from the two independent
loop integrations to decide of the overall factor.
The result is therefore real.

 Other imaginary parts are generated by diagrams of the form 6a and 6c,
in which the insertion of radiative corrections of vertex-type and self-energy
are considered.

Let's now analize Fig.~5d.
In this case - and we shall proceed in a similar way
in the other similar cases -
we prefer to use dimensional regularization to evaluate the sub-cuts which
appear in the relevant Feynman diagrams. The divergences,
in fact, emerge at this level.
The contribution of these sub-diagrams are then folded with a 3-particle cut
integral which is further evaluated by using Sudakov variables in the
Breit frame of the two pion' lines.

The method is particularly convenient
in order to check the infrared safety of the final sum of all the diagrams.

For this purpose let's decompose 5d as shown in 5e, where the top part now
involves a 2-particle cut with 4 on-shell external lines.
$s'=(p_1-k+q_1)^2$, in this particular case,
denotes the invariant energy in the
"s" channel of this subdiagram. In $n=4$ spacetime dimensions this 2-particle
cut is infrared divergent.
The situation is similar to the case discussed above
(diagram 5b), when the two masses of this diagram were set to zero.
Notice that, in principle, also 5b can be regulated dimensionally, although
the expression of its 2-particle contribution is far less obvious,
since it involves an imaginary part with 2 off-shell external lines.

Let's then define the n-dimensional expression of this 2-particle
cut contribution when all the external lines are massless
\beq
\Delta_{box}(s')=\int d^n\,k {\delta_+(k^2)
\delta_+((p_1-k+q_1)^2)\over (2 \pi)^4(p_1-k)^2 (p_2-k)^2}.
\label{gamma11}
\eeq

There exist various possible way to evaluate (\ref{gamma11}), the
simplest probably being the one discussed in \cite{vanner}.
This same method can be simply used to regulate by dimensional regularization
also eq.~\ref{derivation}. In \cite{vanner}, (\ref{gamma11}) is calculated
by the method of Feynman parameters combined with the Kummer series.
A simpler way to obtain this result is, as usual, to simplify the
constraints imposed by the Dirac's delta function in a suitable
frame. The derivation of these discontinuities is quite straightforward
and we omit any of the details and quote the final result.
We get, for the box diagram (top of Fig.~5e)

\beqa
\Delta_{box}(s')&=&
{\pi^{n/2-1}\over 4}{\Gamma[{n\over2}-2]^2\over \Gamma[n/2-1] \Gamma[n-4]}
\nonumber \\
&&\,\,\,\,\,\,\, \times F(1,1,{n\over 2} -1, 1 +{s'\over t}){s'}^{n/2-4},
\label{hyper}
\eeqa
where $F(a,b,c,z)$ denotes, as usual, the hypergeometric
function \cite{bateman}.
The 2-particle cut for the vertex function (Fig. 6b) is instead given
by
\beqa
\Delta_{vertex}&=&\int d^n\,k{ \delta_+(k^2)\delta_+((p_1 -k)^2)\over
(k-p_1)^2} \nonumber \\
&=&-{\pi^{n/2-1}\over 4}{\Gamma[n/2-2]\over \Gamma[n-3]}{s'}^{n/2-3}.
\nonumber \\
\label{vertex}
\eeqa

The expansion around $n=4 +\epsilon$ of eq.~(\ref{hyper}) generates
a single pole in $1/\epsilon$ times a combination of polylogarithmic functions,
the last ones obtained from the Taylor expansion of the
hypergeometric function
\beq
F(1,1,1 + {\epsilon \over 2}, 1+{t\over s'})=
{\left(-t\over s'\right)}^{\epsilon/2-1}\left(1 + {1\over 4}\epsilon^2 Li_2(x)
 +{1\over 8}\epsilon^3(S_{1,2}(x)-Li_3(x)) + O(\epsilon^4)\right),\\
\label{hepexp}
\eeq
where $x={1+t/ s'}$.

The definitions of $Li_2(x)$ and $S_{1,2}$ are given in ref.\cite{BMR}.
 We can easily convolute this result with the expression of the
3-particle cut given in eq.~(\ref{Jzero}) to obtain

\beqa
&&\Delta_{5d}=\int\,d^n\,k\,{\delta_+(k^2)\,\delta_+((p_1-k)^2)
\,\delta_+((p_2-k)^2)}\,\,\Delta_{box}((p_1-k+q_1^2))\nonumber \\
&&=\Gamma_1 \int\,d^n\,k\,{\delta_+(k^2)\,\delta_+((p_1-k)^2)
\,\delta_+((p_2-k)^2)}\nonumber \\
&&\,\,\,\,\,\,\,\times F\left(1,1,n/2-1,1
+t/(p_1-k+q_1)^2\right)\left((p_1-k+q_1)^2
\right)^{n/2-4}\nonumber
\label{boxy}
\eeqa

where we have set
\beq
\Gamma_1 \equiv  {\pi^{n/2-1}\over 2}{\Gamma[n/2-2]^2\over \Gamma[n/2-1]
\Gamma[n-4]}.
\eeq

Notice that all the infrared sensitive parts in (\ref{boxy})
are only contained in the factor $\Gamma_1$.
This is a nice feature of Dimensional Regularization, which allows us to
evaluate completely eq.~(\ref{boxy}) till the last stage.
In fact, defining $\epsilon=n-4$, the pole contribution and the finite
parts of (\ref{boxy}) can be obtained from the relation

\beqa
&& \Delta_{5d}=\mu^{\epsilon}\left(-{\pi\over t}\right)
\left({-t\over \mu^2}\right)^{-\epsilon/2}\Gamma_1
 \int\,d^n\,k\,{\delta_+(k^2)\,\delta_+((p_1-k)^2)
\,\delta_+((p_2-k)^2)\over (p_1-k+q_1)^2},
\label{5d}
\eeqa
where we have introduced a renormalization mass scale $\mu$.
Notice that it is necessary to perform also the remaining $k$ integral
in $4+\epsilon$ dimensions, since it contributes to the finite part.
The basic integral which appears here is
\beqa
&&\sigma_0= \int d^n\,k\delta_+(k^2)
{\delta_+((p_1-k)^2)\delta_+((p_2-k)^2)\over
(p_1-k + q_1)^2}, \\
\eeqa
whose evaluation is similar to the one discussed in Appendix C.
Expanding in $\epsilon$ the logarithmic part of (\ref{5d}) we get
\beq
\Delta^{5d}=\mu^{\epsilon}\Gamma_1 \sigma_0 (1- \epsilon/2
Log[{-t\over\mu^2}]).
\eeq
Inserting the explicit expression of $\sigma_0$ we finally obtain

\beqa
&& \Delta^{5
d}=\mu^{\epsilon}2^{n-5}\pi^{n-5/2}{\Gamma[n/2-2]^2\Gamma[n/2-3/2]\over
\Gamma[n/2-1]\Gamma[n-4]\Gamma[n-3]}F(1/2,3/2,n/2-1,v)\nonumber \\
&&\,\,\,\,\,\,\,\,\, \times {J(Q,s_1,s_2)\over A}\left(1 -{\epsilon\over 2}
Log[{-t\over \mu^2}]\right).
\eeqa

We have defined
\beqa
&&A=s-s_1-2 k^+ q_1^- - 2 k^- q_1^+ \nonumber \\
&&=\,\,\,\,\,\, {(2 Q^2-s_1)(2 Q^2-s_2)(4 Q^4 s - 4 Q^2 s_1 s_2 + s s_1 s_2)
\over (4 Q^4 - s_1 s_2)},
\eeqa

\beqa
 B &=&2 |k_\perp | |q_\perp | \nonumber \\
&=&\,\,\,\,\,\,\left( 2 Q^2 s_1 s_2 (s-2 Q^2)(2 Q^2 s - s_1
s_2)\right)^{1/2}\nonumber \\
&&\,\,\,\, \times {(2 Q^2 -s_1)(2 Q^2 -s_2)\over (4 Q^4 - s_1 s_2)},\nonumber
\\
\eeqa

and set $v\equiv B/A$. Notice that $A\,>\,0,\,$ and $\,B\,>\,0)$.
Notice also that the momenta $k^{\pm}$ are fixed by the (lowest order)
3-particle cut (Fig.~1b)
in terms of $Q$, $s$ and of the two virtualities $s_1$ and $s_2$,
and that $v\approx 1/Q^4\,\,<<\,1$, so an expansion
up to order $v^2$ is a very good approximation to the integrals above.

The evaluation of 6a proceeds in a similar way, though it is slightly more
complex.

In this case we get
\beqa
&&\Delta^{6a}=
\Gamma_2\int d^n\,k \,{\delta_+(k^2)\delta_+((p_1-k)^2)
\delta_+(p_2-k)^2)\over ((p_1-k+q_1)^2)}\Delta_{vertex}((p_1-k+q_1)^2).
\eeqa
Using the explicit expression of eq.~(\ref{vertex}) we get

\beqa
&&\Delta^{6a}=
\Gamma_2\int d^n\,k \,{\delta_+(k^2)\delta_+((p_1-k)^2)
\delta_+(p_2-k)^2)\over ((p_1-k+q_1)^2)^2}\left(1-{\epsilon\over 2}
\,ln\,\,\left({(p_1-k+q_1)^2\over \mu^2}\right)\right). \nonumber \\
\eeqa
We have defined

\beq
\Gamma_2=-{\pi^{n/2-1}\over 2}{\Gamma[n/2-2]\over \Gamma[n-3]}.
\eeq

{}From the $k$ integral we obtain two main terms:
\beq
\sigma_1=\int d^n\,k \,{\delta_+(k^2)\delta_+((p_1-k)^2)
\delta_+(p2-k)^2)\over ((p_1-k+q_1)^2)^2}
\label{sig1}
\eeq

and the logarithmic contribution
\beq
\sigma_2= \int d^4\,k{\delta_+(k^2)\delta_+((p_1-k)^2)\delta_+((p_2-k)^2)\over
(p_1-k + q_1)^2}\,\,ln\left({ (p_1-k+q_1)^2\over \mu^2}\right).
\label{sig2}
\eeq
Their evaluation is discussed in Appendix C and their contribution is
finite. The single pole in $1/\epsilon$,
in diagrams of this type, is generated only by the
2-particle cut of the vertex function shown in Fig. 6b

\beqa
&&\Delta^{6a}={\mu^{\epsilon}\over A^2}J(Q,s_1,s_2)\pi^{n-5/2}
{\Gamma[n/2-2]\over \Gamma[n-3]\Gamma[n/2-3/2]}\nonumber \\
&&\,\,\,\,\,\times\left( 2^{n-5}{\Gamma[n/2-3/2]^2\over \Gamma[n-3]}\,\,
F(1,3/2,n/2-1,v)
(1-\, Log[{(1+v)A\over \mu^2}]) \right.\nonumber \\
&&\,\,\,\,\,\,\,\, \left. -\pi\,\epsilon \,{v\over 1+v}\right)
\eeqa

Again, in practical applications, just the first few terms in $v$ are
necessary.

A radiative correction of another type, also
 responsible of the generation of imaginary parts is shown
in Fig.~(6c), and it corresponds to a self energy insertion on the fermionic
line at the top of the box diagram. An odd number of particle cuts is
involved in this diagram. As we have discussed in the former cases,
also here, again, we can unfold the 5-particle cut into a 2- times
a 3-particle cut. Both cuts are regular in $n=4$ dimension, so there is no
need to use dimensional regularization.
For the 2-particle cut we get ($q$ is the external momentum)
\beq
\int d^4 k\,\delta_+(k^2)\,\delta_+((q-k)^2)={\pi\over 2}\theta(q^2),
\eeq
and the 5-particle cut can then be set in the form
\beqa
&& \int d^n\,k\,{\delta_+(k^2)\delta_+((p_1-k)^2)\delta_+((p_2-k)^2)
\theta ((p_1-k+q_1)^2)\over \left((p_1-k+q_1)^2\right)}.\nonumber \\
\label{theta}
\eeqa
It is not so difficult to show that the condition $(p_1-k+q_1)^2\,\,>\,\,0$
is identically satisfied for all possible values of $k$ in the integral.

Evaluating this integral at $n=4$ we get

\beqa
\Delta^{6d}=-{\pi\over 2 s t A (1-v^2)}.
\eeqa
The evaluations presented in this section, as we have seen, can be
carried on up to the last stage. This is a remakable feature.
 A simple factorized structure
for the imaginary parts generated by the radiative corrections to the lowest
order sum rules emerges. All these corrections, (see Fig.~7), can be
easily evaluated by these methods.

It is not difficult to realize that the set displayed in Fig.~7 is
the complete one. The diagrams shown in Fig.~8, for instance, have not been
added to the list. Let's see why.

As an example, let's consider the diagram shown
in Fig.~8a. From a first inspection, it seems that this diagram should
be included to the list since the number of its cuts is odd.
However,  it is easy to realize that it involves a subcut of $t$-type,
which is non vanishing only if $t$ is in the unphysical region ($t\,\,>\,\,0$).
Therefore 8a can not contribute as far as $t$ remains in the physics region.

Other contribution to the real part of the spectral density are dispalyed
in Figs.~8b and 8c. Each of these two diagrams has a maximal number of
internal on-shell lines and no other additional line can go on shell.
In fact any other additional line, if on-shell, induces at least one vertex
with 3 on-shell massless lines in the new diagram. The presence of any vertex
of this type then, will force all the 3 particles at that vertex
to be collinear. This condition is, in general, too strong to give a
non vanishing discontinuity, except at specific points in the phase space
of the process. The discussion presented in section 3 may serve well illustrate
this aspects. In fact, in that case,
$t$ was forced to be zero in the presence of such configurations
in the reduced diagram.

\section{Conclusions}
We have discussed in detail some crucial aspects
concerning the analyticity properties of the spectral densities
in Compton processes.
The analytical bounds on the region in which a dispersion relation
is valid is controlled by the quark mass in the expansion of the correlator.

For nonvanishing $m$ the spectral density has special singularities,
absent in the form factor case, which have been identified as Landau
surface of $u$-type. At order $\alpha_s$ they are important
in the analysis of the radiative corrections.

As we have seen, in the massless case, a simpified
picture of such corrections emerges. Remarkably, the evaluation of the
complex parts, responsible of effect of interference in the sum rules,
can be done in close form. For this purpose,
Sudakov methods in $n$ dimensions and
dimensional regularizations have been developed and combined in an original
way.
Since it is generally believed that,
at intermediate energy,
the phases of elastic scatterings are of non-perturbative origins,
it is interesting to see how relevant the "Feynman mechanism" \cite{NR1}
which motivates
the sum rule description is - compared to
 hard scattering factorization -  in the description of such effects.

We reserve to discuss the
phenomenological application of our results to the evaluation of the phases
of pion photoproduction elsewhere\cite{cchn}.

However, from this work, a well defined structure for the complex
spectral densities for $\gamma$ $\gamma$ collisions emerges, which can be
enforced in an actual calculation.

\vbox{\vskip 0.3 true in}
\centerline{Acknowledgements}
I am very grateful to Yuri Dokshitzer for endless discussions
and suggestions
on the matter presented, which have made this work possible, and to
Hsiang-nan Li, George Sterman and Alan White
for many suggestions.
I warmly thank Rajesh R. Parwani for all the help
and the advice provided on this matter
and E. Berger, L. Bergstr\"{o}m, G.T. Bodwin, T.H. Hansson,
R. Liotta, H.R. Rubinstein and C. Zachos for illuminating discussions.

Finally, I am grateful to the Theory Depts. of Stockholm and Lund University,
the Academia Sinica of Taiwan and to the Theory Group at Saclay
for their kind hospitality.

\renewcommand{\theequation}{A.\arabic{equation}}
\setcounter{equation}{0}
\vskip 1cm \noindent
\noindent {\large\bf Appendix A. The light-cone formulation. }
\vskip 3mm \noindent

Here we summarize briefly the kinematics of Compton processes
and point out the
importance to work out the expression for the spectral densities in the
light-cone frame since, in such a frame, they become polynomial.
The simplification is enormous since the Mandelstam function
(see eq.~\ref{trian})
-which appears in the form factor case- is absorbed into the definition of a
large "plus" ligh-cone momentum of one of the off shell parton' lines.
It is important to notice that, since from the very beginning we have been
working with a diagrammatic expansion of current correlators,
in each diagram in the O.P.E. (see Fig.~1) the virtuality of
the pion' lines will, in general, be different.

Nevertheless we can picture Fig.~1 as an ordinary scattering
process and select a special frame in which all the transversal
momentum belongs to just a single particle (the photons).
Let's define

\beq
s=(p_1+q_1)^2        \,\,\,\,\,\,    t=(q_2-q_1)^2
         ,
u=(p_2-q_1)^2, \eeq
with

\beq
s+t+u=s_1+s_2.
\eeq
%We consider both $s$ and $t$ to be very large and in the physical region
%($s>0,t<0$ ).

We also expand the momenta of the incoming photon as
\beqa
q_1 &=& q_1^+ n^+ +q_1^- n^- +q_{1\perp}.
\eeqa
In the Breit frame of the incoming meson we have
\beq
u=(p_2-q_1)^2=2 Q^2-s +{{s_1 s_2}\over 2 Q^2}.
\eeq

Covariant expressions for $q_1^\pm$ and $q_2^\pm$ can be easily
obtained in the form

\beq
q_1^+={(s-2 Q^2)(2 Q^2-s_2)\over 2Q\delta},
\eeq

\beq
q_1^-={(2 Q^2-s_1)(2 Q^2 s-s_1 s_2)\over 4Q^3\delta}.
\eeq
%%%%%%%%%%%%%%%%%%%%%%%%%

In the light cone frame the spectral function
can be calculated to lowest order
in terms of 3-cut integrals of the form \cite{CRS} \cite{co}
\beq
\label{if}
I[f(k^2,k\cdot p_1,...)]=\int
d^4k   f(k^2,k\cdot
p_1,...){\delta_+(k^2)\delta_+((p_1-k)^2)
\delta_+((p_2-k)^2)\over (p_1-k + q_1)^2},
\eeq
\\
\beq
\label{ipf}
I'[f(k^2,k\cdot
p_1,...)]=\int d^4k
f(k^2,k\cdot p_1,...)\delta_+(k^2)\delta_+((p_1-k)^2)\delta_+((p_2-k)^2), \eeq
and as in standard evaluation of Feynman diagrams, we can reduce tensor
discontinuities to scalar one by any reduction procedure.

Mass dependent integrals can be analized quite easily in the light cone frame.

For illustrative purposes consider
\beqa
 & & J_{34}(s_1,s_2,s,t,m_1,m_2) \nonumber \\
& & = \int d^4 k {\delta_+(k^2)\delta_+((p_2-k)^2 -
m_2^2)\delta_+((p_1-k)^2-m_1^2)
\over (p_1 - k + q_1)^2}.
\label{disc}
\eeqa
In this integral, the $k^{\pm}$ components of the internal momentum are fixed
at the values
\beq
k^+={{q \left( -2 {  m_1} {Q^2} + 2 {  m_2} {Q^2} +
       {  m_1} {   s_2} - 2 {Q^2} {   s_2} +
       {   s_1} {   s_2} \right) }\over
   {-4 {Q^4} + {   s_1} {   s_2}}},
\eeq

\beq
k^-={{q \left( 2 {  m_1} {Q^2} - {  m_1} {   s_1} +
       {  m_2} {   s_1} + 2 {Q^2} {   s_1} -
       {   s_1} {   s_2} \right) }\over
   {4 {Q^4} - {   s_1} {   s_2}}}.
\eeq
We get
\beq
J_{34}(s_1,s_2,s,t)=(-2\pi i)^3{Q^2\over 2 (4 Q^4-s_1 s_2)} T_{ang}
\eeq

with

\beqa
T_{ang}&=&\int_{0}^{2 \pi}{d\theta  \over A +
 B  cos[\theta]} \nonumber \\
&& = 2 {\pi \over {\left( A_2^2 - B_2^2\right)}^{1/2}}
\eeqa
\beqa
A_2&=& s + m_1 -2 k^+ p_1^- -2 k^- p_1^+ -2 k^+ q_1^- -2 k^- q_1^+; \nonumber
\\
B_2&=&{\left ( 8 q_1^+ q_1^- (2 k^+ k^- -m_1) \right )}^{1/2}.
\eeqa
Remarkably $T_{ang}$ is a rational function of $Q^2(s_1,s_2,t)$ in the
Breit frame even for nonvanishing $m_1$ and $m_2$. This is due to the identity

\beq
A^2 - B^2 ={w(m_i,Q^2,s_1,s_2)\over
  \left( -4 {Q^4} + {   s_1} {   s_2} \right)^2} \nonumber
\eeq

\beqa
w(m_i,Q^2,s_1,s_2)&=& \left( -4 {  m_1} {Q^4} + 2 {  m_2} {Q^2} s +
         4 {Q^4} s + 2 {  m_1} {Q^2} {   s_1} \right.\nonumber \\
&& \left. -
         2 {  m_2} {Q^2} {   s_1} -
         2 {Q^2} s {   s_1} -
  2 {  m_1} {Q^2} {   s_2} -
         2 {Q^2} s {   s_2}\right. \nonumber \\
&&\left.  -
         {  m_1} {   s_1} {   s_2} + s {   s_1} {   s_2} \
         \right)^2
\eeqa

which gives
\beqa
J_{34}(s_1,s_2,s,t)&=& {\pi \over 2 \delta W(m_1,m_2)},
\eeqa
and is the result given in eq.~(\ref{w}).

\renewcommand{\theequation}{B.\arabic{equation}}
\setcounter{equation}{0}
\vskip 1cm \noindent
\noindent {\large\bf Appendix B. Radiative Corrections }
\vskip 3mm \noindent
The evaluation of eq.~(\ref{almostfinal}) proceeds as follows.
The integrations in the $l^-$, $l_\perp^2$,
and in the angular variable $\phi$ in the
transversal plane can be done explicitely. We define
\beq
cos\, \phi={|q_\perp\cdot l_\perp|\over |q_\perp||l_\perp|},
\eeq
and since $q_{1\perp}=q_{2\perp}$, we obtain the relation
$q_{12\perp}\cdot l=0$.
The only angular dependence in $\phi$, in the integral, is
contained in the factor $(q_1-l)^2$.
Therefore we obtain
\beqa
&& I_{5b}=-{1\over 4 q_{12}^+}\int_{a0}^{a1}dl^+\,\,
\int_{0}^{2 \pi} \d\phi{\Delta_2(l^+,l_m)\over
((q_1^+ l_m + q_1^- l^+ -2 |q_\perp||l_\perp |cos\phi)}\nonumber \\
&& ={\pi\over 4 q_{12}^+}\int_{a0}^{a1}dl^+ {\Delta_2(l^p,l_m)\over
\left( (q_1^+ l^- + q_1^- l^+)^2 - 4 q_1^+ q_1^- l^+ l_m \right)^{1/2}},
\nonumber
\label{derivation1}
\eeqa
which gives, after integration, eq.~(\ref{derivation}).

$\mu_1$ and $\mu_2$ are momenta at the boundary,
fixed by the conditions on the energy flow.
In order to fix these two momenta let's rewrite these conditions in the
light cone variables

\beqa
&& l^+ + l_m > 0,  \nonumber \\
&& q_1^+ +l^+ +q_1^- +l^- >0, \nonumber \\
&& q_{12}^+ +l^+ +q_{12}^- + l^- >0, \nonumber \\
&& s'=(p_1-l)^2=s_1 -2 p_1^+ l^- -2 p_1^- l^+ >0.
\label{constr}
\eeqa
In the last condition above,
which is simply the requirement of $s'>0$,
 we have for simplicity neglected
the contributions coming from the regulator mass $\mu$ and from $m$
(specifically $(s'\,\,>\,\,(m +\,\mu)^2)$).
 This approximation is not going to affect our conclusions in
 any significant way.

The analysis of these conditions is quite involved.
Here we summarize the basic results.

In the Breit frame these conditions are polynomial constraints
in the variable $Q^2$ and can be re-stated respectively in the form
\beqa
l^+\,\,>\,\,\mu_1, \nonumber \\
l^+\,\,>\,\,\mu_2, \nonumber \\
l^+\,\,>\,\,\mu_3, \nonumber \\
l^+\,\,<\,\,\mu_4, \nonumber \\
\eeqa
where
\beqa
&& \mu_1={(-2 Q^2 + s_1)(-2 Q^2 + s_2)\over
2 Q (4 Q^4 - s_1 s_2)(-4 Q^4 + s_1 s_2)^2(16 Q^8 -16 Q^6 s + 8 Q^4 s^2
-4 Q^2 s s_1 s_2 + s_1^2 s_2^2)},\nonumber \\
\eeqa

\beqa
&& \mu_2= \nonumber \\
&& {Q(s - 2Q^2)(-2 Q^2 +s_2)(-16 Q^6 + 8 Q^4 s +4 Q^4 s_1 -
4 Q^2 s s_1 +4 Q^4 s_2 + s_1^2 s_2 - s_1 s_2^2)\over
(-4 Q^2 + s_1+ s_2)(-4 Q^4 + s_1 s_2)^2},\nonumber \\
\eeqa

\beqa
\mu_3={(-2 Q^2 + s_2)P_3(Q,s_1,s_2,s,t)\over 2 Q (4 Q^2 -s_1 -s_2)
(-4 Q^4 + s_1 s_2)^2},
\eeqa

and where
\beqa
P_3(Q,s_1,s_2,s,t)&=&(-32 Q^{10}+32 Q^8 s -16 Q^6 s^2 -16 Q^6 s s_1 +
8 Q^4 s^2 s_1 \nonumber \\
&& \,\,\,\,\,+16 Q^8 s_2 +8 Q^4 s s_1 s_2 +8 Q^4 s_1^2 s_2 -
4 Q^2 s s_1^2 s_2 \nonumber \\
&& \,\,\,\,\,\, - 8 Q^4 s_1 s_2^2 - 2 Q^2 s_1^2 s_2^2 +
s_1^2 s_2^3),\nonumber  \\
\label{p3}
\eeqa

\beqa
 \mu_4 &=& 2 Q^3 (-2 Q^2 + s_2)(16 Q^8 -16 Q^6 s + 8 Q^4 s^2 +8 Q^4 s s_1 -
\nonumber \\
&& \,\,\,\,\,\,4 Q^2 s^2 s_1 -4 Q^2 s s_1 s_2 -4 Q^2 s_1^2 s_2
+2 s s_1^2 s_2 + s_1^2 s_2^2).
\eeqa

It is possible to prove the validity of the following inequalities

\beq
0\,\,<\,\,\mu_2\,\, <\,\, \mu_3\,\, <\,\,\mu_1\,\, <l^+ \,\,<\mu_4
\eeq
and hence the integration over the variable $l^+$
in (\ref{derivation1}) is cut from below by $\mu_1$ and from above by $\mu_4$.

\renewcommand{\theequation}{C.\arabic{equation}}
\setcounter{equation}{0}
\vskip 1cm \noindent
\noindent {\large\bf Appendix C. Sudakov methods in $n$ spacetime dimensions }
\vskip 3mm \noindent

In this appendix we cover some of the technical derivation of the
$n$ dimensional discontinuities, obtained by using a special technique
which is essentially based on the use of $n-2$ transversal dimensions
for the Sudakov variables in the Breit frame \cite{pepevafa}.

The evaluation of the longitudinal parts of the integrals proceeds in the same
way as illustrated in Appendix B. Therefore the $k^{\pm}$ components of the
momenta inside each loop of integration are fixed. The evaluation of these
longitudinal parts brings in a jacobean, which is polynomial in the
momentum $Q$, which is evaluated on the two dimensional subspace of
$n^{\pm}$. The transversal integral, instead, is expressed in terms of
$n-2$ angular variables. Of these, $n-3$ can be integrated trivially,
while the remaining one, which we call $\theta$, is non trivial.
The angular integral so generated are finite, but we need there expression
up to $O(\epsilon)$, since also these terms
contribute to the finite part of the spectral densities.
We illustrate here the derivation of $\sigma_1$ and $\sigma_2$, as defined
by eqs.~(\ref{sig1}) and (\ref{sig2}).

We obtain
\beq
\sigma_1 ={1\over A^2}J(Q,s_1,s_2) \omega(n) T_{ang,1}(n,v)
\eeq

where
\beq
J(Q,s_1,s_2)={Q^2\over 2 (4 Q^4 - s_1 s_2)}
\eeq
is the two-dimensional jacobean, while
\beq
\omega(n)=2 {\pi^{n/2-3/2}\over \Gamma[n/2-3/2]}
\eeq
is a volume factor obtained from integration over $n-3$ angular dimensions.
The nontrivial part is contained in
\beq
T_{ang,1}(n,v)=\int_{0}^{2 \pi}d\,\theta
{sin\,\,\theta^{n-4}\over (1 + v cos\,\,\theta)^2},
\eeq
which is of hypergeometric form

\beq
T_{ang,1}(n,v)={2^{n-5}\over (1+v)^2}{\Gamma[n/2-3/2]^2\over \Gamma[n-3]}
F(2,n/2-3/2,n-3, {2 v\over 1 +v}).
\label{tango}
\eeq

Using the quadratic (Gauss) relation
\beq
F(a,b,2 b,z)=(1- {z\over 2})^{-a} F(1/2 a,a/2 +1/2, b+1/2,
\left({z\over 2 -z}\right)^2)
\eeq
we can re-express (\ref{tango}) in the form
\beq
T_{ang}(n,v)= 2^{n-5}
{\Gamma[n/2-3/2]^2\over \Gamma[n-3]}F(1,3/2,n/2-1,v).
\label{tango1}
\eeq
Notice that, being $v\,<\,<\,1$, the hypergeometric function in (\ref{tango1})
can be approximated, for all the practical purposes, by the first 2 terms,
namely up to order $v$.

$\sigma_2$ can not be reduced to a complete explicit expression,
because of the logarithmic corrections, which are not exactly
of the hypergeometric form. This contribution can be directly evaluated with
$n$ set to be 4 from the beginning, since these corrections
already appear with a factor of $\epsilon$ away from $n=4$.

For this second integral, after that the longitudinal
integration has been performed, we obtain

\beq
\sigma_2={J(Q^2,s_1,s_2)\over A^2} \omega(n) T_{ang,2}(v,n),
\eeq

where

\beqa
T_{ang,2}&\equiv &\int_{0}^{\pi}d\,\theta sin\,\,\theta ^{n-4}
{Log [(A + B cos\,\,\theta)/\mu^2]\over (1 + v cos\,\,\theta)^2}
\nonumber \\
&&= Log [{A\over \mu^2}] T_{ang,1}(n,v) + T_{ang,3}(n,v).\nonumber \\
\eeqa
We have defined
\beq
T_{ang,3}\equiv {2^{n-5}\over (1+v)^2}\int_{0}^{1}d\,t
Log[(1+v)(1- z\, t)]t^{n/2-3/2}(1-t)^{n/2-5/2}(1-z\, t)^{-2},
\eeq
 and $z=2 v/(1+v)\approx 2\,v$.
Expanding the logarithmic part to lowest order as
$Log[1-z\,t]\approx -z\,t + O(v^2)$
and taking $n=4$, we get
\beq
T_{ang,3}\approx Log[1+v]\,\,T_{ang,1}(4,v)
+ {2 v\over (1+v)}B[3/2,1/2] +\,\,\,\, O(v^2)
\eeq
Combining all the terms together we finally obtain
\beq
T_{ang,2}=Log[{A(1+v)\over \mu^2}]\,\,T_{ang,1}(4,v)
+ {\pi v\over (1+v)}
+\,\,\,\, O(v^2).
\eeq

Using these results for the angular integrals, it is straightforward to
obtain the explici expressions of $\Delta^{5d}$ and $\Delta^{6a}$.

\renewcommand{\theequation}{D.\arabic{equation}}
\setcounter{equation}{0}
\vskip 1cm \noindent
\noindent {\large\bf Appendix D. Borel methods }
\vskip 3mm \noindent

In this appendix we present a discussion of how and of when the Borel method
\cite{NR1} can be applied to Compton scattering.
We have organized the material starting from 2 point functions. Then
we come to the pion form factor \cite{NR1}.
A simple way to invert Borel transformed amplitudes for 3-point functions
is presented. Then we show that the method can be extended to lowest order
Compton scattering. In this case it is
shown that the divergence of the Schwinger
representation in the box diagram can be kept under control by working
in the euclidean region.

To make our discussion self-contained we first illustrate briefly
the simplest application of this
technique \cite{NR1} to correlators of lower rank.

Let's consider the dispersion relation  of the polarization operator
\beq
\Pi(q_1^2)={1\over \pi}\int_{0}^{\infty} {
\,\Delta(s)ds\over (s-q_1^2)}
 \,\,\, + subtractions,
\label{pd}
\eeq
 with a singularity cut starting at $q_1^2=0$.
Eq. (\ref{pd}) can also be written in the form
\beqa
\Pi(q_1^2) &=&{1\over\pi}\int_{0}^{\infty}ds\int_{0}^{\infty}d\alpha
\,\Delta(s) e^{-\alpha(s-q_1^2)}, \,\,\,\,\ q_1^2<0,
\label{pd1}
\eeqa
where we have used the exponential parametrization of the
denominator.
%%%%%%%%%%%%%%%%%%%%

The Borel transform in one variable is defined in its
differential version by the operator \cite{SVZ}
\beq
B(Q^2\to M^2)=lim_{\stackrel{Q^2,n\to \infty}{Q^2/n=M^2}}\frac{1}{(n-1)!}
(Q^2)^n (-\frac{d}{dQ^2})^n.
\label{dif}
\eeq
$M^2$ denotes the Borel mass.

It satisfies the identity
\beq
B(Q^2\to M^2) e^{-\alpha Q^2}=\delta(1-\alpha M^2) \,\,\,\,\ \alpha Q^2>0.
\label{boq}
\eeq

Acting on the polarization operator $\Pi(q_1^2)$
with the Borel transform we get the usual exponential suppression of the higher
states
\beq
M^2 B(-q_1^2\to M^2)\Pi(q_1^2)={1\over \pi}\int_{0}^{\infty}ds
\,\Delta(s)e^{-s/M^2}.
\label{MBP}
\eeq
At this point we can Borel transform once again eq.~(\ref{MBP}),
with respect to the inverse Borel mass $1/M^2$, in order to obtain
\beq
B(1/M^2\to \nu)( M^2 B(-q_1^2\to M^2)\Pi(q_1^2))=\frac{1}{\nu}\Delta(1/\nu).
\label{BMMP}
\eeq
Therefore, by acting iteratively with Borel transforms
on $\Pi(q_1^2)$, we obtain an expression
from which we can
 easily identify the spectral weight $\Delta(s)$ of eq.~(\ref{pd}).

Let's now turn our attention to vertex functions (see also Fig. 1c).

The Borel transformed amplitude for the pion form factor
(the complete spectral density) is
given by \cite{NR1}

\[
\phi(M_1^2,M_2^2,q^2)={1\over \pi^2}\int_{0}^{\infty}ds_1\int_{0}^{\infty}ds_2
\,\rho_{\pi 3}^{pert}(s_1,s_2,q^2)\, e^{-s_1/M_1^2 - s_2/M_2^2}
\]
\beqa
&& ={3\over 2 \pi^2 (M_1^2 + M_2^2)}\int_{0}^{1} dx\, x (1-x)
\,exp\left({-x q^2\over(1-x)(M_1^2 + M_2^2}\right).
\label{fi}
\eeqa
To isolate the spectral function from
 eq.~(\ref{fi}) we need to use Borel transforms
and act on it with the differential operator
\beq
\label{BO}
B(1/M_1^2\to 1/\nu_1)B(1/M_2^2\to 1/\nu_2) M_1^2 M_2^2.
\eeq
Here we discuss a possible way of doing this.
The inversion of (\ref{fi}) can be obtained by taking
inverse Lapace transforms twice
- respect to $1/M_i^2$ - of this equation.
In fact, for a given function $F(M^2)$, the following identity
\beq
 {L}^{-1}(1/M^2\to\nu)F(M^2)=(1/\nu) B(1/M^2\to 1/\nu)F(M^2)
\eeq
relates the differential operator given in (\ref{dif}) to the inverse Laplace
transform
\beq
 {L}^{-1}(1/M^2\to\nu)={1\over 2 \pi i}
\int_{c-i\infty}^{c+ i \infty}d(1/M^2)\, exp\left ({\nu/M^2} \right ).
\label{inv}
\eeq
Defining $1/M_1^2= \mu_1, 1/M_2^2 =\mu_2$, then , we are
required to act with the operator defined in eq.~(\ref{BO}) on the integral
function
\beq
\chi(\mu_1,\mu_2,q^2)={1\over \mu_1 + \mu_2} \int_{0}^{\infty}dx\, { x
\over (x+1)^4}\,exp\left(-x Q^2 \mu_2 + {x q^2\mu_2^2 \over \mu_1 +
\mu_2}\right).
\label{chi}
\eeq

Using the gaussian relation
\beq
exp\left(\alpha^2/(4 k)\right)=\left(k/\pi\right)^{1/2}\int_{-\infty}^{\infty}
d\sigma \,exp({-k
\sigma^2 - \alpha \sigma} )
\label{gau}
\eeq
we can rewrite eq. (\ref{chi}) into the form

\beqa
&& \chi(\mu_1,\mu_2,q^2)  \nonumber \\
&& =\int_{-\infty}^{\infty}
d\sigma \int_{0}^{\infty}  dx {x\,exp\left({-x q^2 \sigma - (\mu_1 +\mu_2)
\sigma^2 -2 x^{1/2} q \mu_2 \sigma}\right)
\over (x+1)^4(\pi\, (\mu_1 + \mu_2))^{1/2}}.
\label{chimu}
\eeqa
By using the relation
\beq
 {L}^{-1}(\mu_1\to\nu_1) {1\over(\mu_1 +\mu_2)^{1/2}}exp\left(-\mu_1 +\mu_2
\sigma^2\right)
=\left( {\pi\over(\nu_1-\sigma^2)}\right)^{1/2} \Theta(\nu_1-\sigma^2)
\label{inv1}
\eeq
on (\ref{chimu}) we finally get
\beqa
&& {L}^{-1}(\mu_2\to\nu_2) {L^{-1}}(\mu_1\to\nu_1) \chi(\mu_1,\mu_2,q^2)
\nonumber \\
&& =  {L^{-1}}(\mu_2,\nu_2) \int_{0}^{\infty}{x dx\over (x+1)^4}
\int_{-\sqrt{\nu_1}}^{\sqrt{\nu_1}} d\sigma
 {exp\left ( (-x q^2 \mu_2 -2 x^{1/2}q \mu_2 \sigma -\mu_2)
\nu_1\right )\over (\nu_1-\sigma^2)^{1/2}}  \nonumber \\
&&  \quad \quad \quad  = \int_{\sqrt{\nu_1}}^{\sqrt{\nu_1}}d\sigma{(-\sigma +
f^{1/2})
^3 q^4\over \left(\sigma - f^{1/2})^2 + q^2 \right)^4  f^{1/2} (\nu_1-
\sigma^2)^{1/2}}\nonumber  \\
&& \quad \quad \quad = (1/6) q^4 (d/dq^2)^3 Y(q^2,\nu_1,\nu_2),
\label{inv2}
\eeqa
where
\[
Y(q^2,\nu_1,\nu_2)=\int_{-\sqrt{\nu_1}}^{\sqrt{\nu_1}} d\sigma {(\sigma -
f^{1/2})^3\over f^{1/2}(\nu_1- \sigma^2)^{1/2} \left((\sigma -f^{1/2})^2 +
q^2\right)
}, \]

\beq
f=\sigma^2 - \nu_1 -\nu_2.
\label{Yq}
\eeq
By redefining $\nu_{i}=s_{i}, \, i=1,2$, and after doing the explicit
integration
 of in eq.~(\ref{Yq}),  it is easy to relate this last result
to the 3-particle cut integral for
the triangle diagram   (Fig. 1b), in the scalar case,
\beq
\label{Yoq}
{Y(q^2, s_1,s_2)\over 4 q^2} =\Delta_3 (s_1,s_2,t)
\eeq
 The derivative with respect to $q_1^2$ in eq. (\ref{inv2})
 takes into account
the fermionic character
of the propagators in Fig. 1b compared to the scalar case
(see eq. (\ref{Yoq})).
 The spectral function for the pion form factor
can then be expressed in the form \cite{NR1}
\beqa
&&{\rho_\pi}^{pert}_3(s_1,s_2,t) \nonumber \\
&& \quad \quad =
{3\over 2 \pi^2} t^2 \biggl( \left ({d\over dt}\right )^2 +
{t\over 3} \left ({d\over dt}\right )^3 \biggr)
{1\over ((s_1 + s_2 - t)^2 - 4 s_1 s_2)^{1/2}}
\label{ropi}
\eeqa

with $t=-q^2$.

For four-point functions
this procedure simplifies considerably. The method can be applied
exactly as in
the form factor case, although it is necessary to work in the euclidean
region from the beginning.
However 1) the amplitude has $first$
to be continued in the unphysical $s$ and $t$ region
before any Borel transform is taken
2) afterwards the result has to be continued back to the physical region.
These two steps allow us to bypass problems related to the divergences of the
Schwinger representation (see the discussion below) of the amplitude.
Since, by doing so, we are comparing $timelike$ and $spacelike$
representations of the same amplitude, we analize step by step the
effect of the Borel transforms $both$ on the timelike amplitude and on the
spacelieke one.
As we have already emphasized, the method
works $only$ if the dispersion relation is unbound, i.e. the dispersion
integral
can be extended to infinity for both $s_1$ and $s_2$.
We have pointed out that for massless correlators this is indeed the case.

Let's work in the euclidean region of $T_4(p_1^2,p_2^2,s,t)$
with
spacelike external invariants ($p_1^2,p_2^2<0$,
$s=(p_1+q_1)^2<0$, $t=(q_2-q_1)^2<0$).
It is then possible to relate $T_4$ to its euclidean continuation $T_{4E}$
and use the Schwinger parametrization for the latter:

\beq
T_{4E}=\int d^4k\,\int_{0}^{\infty} [d\alpha_i]\,exp\left ({-\alpha k^2 - \beta
(p'_1-k)^2
-\gamma (p'_1-k+q'_1)^2 -\epsilon (p'_2-k)^2}\right ),
\label{ex}
\eeq
where $\alpha_i$ is a short notation for the proper time parameters.
Divergences, in this representation, reappear when we move into the physical
$s,t$ region.
We perform the integration over the loop momentum in eq. (\ref{ex}) to get
\beq
T_{4E}=\int_{0}^{1}dx_1dx_2 dx_3 dx_4\delta(1-x_1-x_2-x_3-x_4)
\int_{0}^{\infty}\Sigma^3 d\Sigma e^{\tau}
\label{ex1}
\eeq
where
\beq
\tau=-\Sigma(A_1(x_i) s_1 -A_2(x_i) s_2 - A_3(x_i,t))
\eeq
with suitable expression for $A_1,A_2,A_3$.
In the region where $p_1^2=-{p'}_1^2$
$p_2^2=-{p'}_2^2$ and, in general, $p_i\cdot p_j=-{p'}_i\cdot {p'}_j$,
$T_{4E}$ is related to $T_4$
by an analytic continuation
\beq
T_4({p}_1^2,{p}_2^2,s,t)=i T_{4E}({p'}_1^2,{p'}_2^2,s',t')
\label{eu}
\eeq

In the euclidean region the form of the dispersion relation becomes
\beq
T_{4E} =-{i\over 4 \pi^2}\int_{0}^{\infty}ds'_1\int_{0}^{\infty}ds'_2
{\Delta
(-s'_1,-s'_2,s',t')\over (s'_1 +{p'_1}^2)(s'_2 +{p'_2}^2)}\,\,\, +\,\,\,\,
subtr.
\label{exte}
\eeq
Exactly as in the case of the polarization operator (eqs. (\ref{pd}) and
(\ref{pd1})),
we can now apply Borel transforms on eq.~(\ref{exte}) to get
 \beqa
\eta_E(M_1^2,M_2^2,s',t')\nonumber \\
&&= B({p'_1}^2\to 1/M_1^2)\,B({p'_2}^2\to
1/M_2^2)\,T_{4E}({p'}_1^2,{p'}_2^2,s',t')  \nonumber \\
&=& -{i\over 4\pi^2}\int_{0}^{{\infty}}ds'_1\int_{0}^{{\infty}}ds'_2
e^{-s'_1/M_1^2
-s'_2/M_2^2} \Delta(-s'_1,-s'_2,s,t) \nonumber  \\
 &&= \pi^2\int_{0}^{1}dx_1\,dx_2\,dx_3\,\delta(1-A_1 \Sigma M_1^2)\delta(1-A_2
M_2^2\Sigma ) e^{-\Sigma A_3} \Sigma^3 \,d\Sigma\nonumber \\
&&=\pi^2 \int_{0}^{1}{ dx_1 \,dx_2 \pi^2\over M_1^2
M_2^2 \,x_1^2 \,x_2}e^{-b_1/M_1^2 -b_2/M_2^2} \\
\label{eta}
\eeqa
where we have defined
\beqa
b_1 &=& s {(1-x_1-x_2)\over x_2}\nonumber  \\
b_2 &=& {(-s x_1 + t x_2)\over x_1}.
\label{bi}
\eeqa
Notice that $B({p'_i}^2\to M_i^2)=B(-p_i^2\to M_i^2)$, since $p_i^2<0$
by assumption.

By acting similarly on $T_4$  as given by (\ref{ext})
- with $B(-p_i^2\to 1/M_i^2)$ -
we also get
\beqa
\eta (M_1^2,M_2^2,s,t) &=&-{1\over 4
\pi^2}\int_{0}^{\infty}ds_1\int_{0}^{\infty}ds_2\,
\Delta(s_1,s_2,s,t)e^{-s_1/M_1^2-s_2/M_2^2} \nonumber  \\
 &= & B(-p_1^2\to M_1^2)\,B(-p_2^2\to M_2^2)\,T_4 (p_1^2,p_2^2,s,t).
\label{eta1}
\eeqa

 Reapplying the Borel transform (this time with respect to $1/M_i^2$)
sequentially  on both $\eta$ and $\eta_E$ and using eq. (\ref{boq})
 we get respectively
\[
B(1/M_1^2\to \nu_1)\,B(1/M_2^2\to \nu_2)\,(M_1^2 M_2^2\,
\eta_E(M_1^2,M_2^2,s',t'))
\]
\beqa
 & =&\int_{0}^{1}\int_{0}^{1} {dx_1 dx_2\over x_1^2 x_2} \delta(1-b_1 \nu_1)
\delta(1-b_2\nu_2) \nonumber \\
 &=&-{1\over 4 \pi^2 \nu_1\nu_2}\Delta_E(-1/\nu_1,-1/\nu_2,s',t')
\label{uno}
\eeqa
for the euclidean part,and
\[
B(1/M_1^2\to\nu_1)\,B(1/M_2^2\to\nu_2)\,(M_1^2 M_2^2\,\eta(M_1^2,M_2^2,s,t))
\]
\beq
=-{1\over 4 \pi^2 \nu_1\nu_2}\Delta(1/\nu_1,1/\nu_2,s,t).
\label{due}
\eeq
in the physical region.
As we can see from eq. (\ref{uno}), the integral on the two parameters $x_1$
and $x_2$
 is now trivial and can be performed straightforwardly.
A simple calculation then gives
\beq
\Delta_E(-1/\nu_1,-1/\nu_2,s't')={4\pi^4\over{\nu_1 \nu_2 s' t'}}.
\label{treE}
\eeq
Using the identification $1/\nu_1\to s_1, 1/\nu_2\to s_2$,
analytically continuing back (the euclidean result in (\ref{treE})
with the prescriptions $s'\to -s$ $t'\to -t$
and using eq. (\ref{eu}) we finally get
\beq
\Delta(p_1^2,p_2^2,s,t)=-4i{\pi^4\over s t},
\eeq
valid in the physical region.
This result agrees with the one obtained from the cutting rules eq.
(\ref{set}).
The Borel method allows
to overcome, in this simpler case,
the difficulties related to the evaluation of discontinuity
integrals. The original discussion of this section
therefore provides an extension of the method
to Compton scattering, from the known case of the form factor \cite{NR1}.
 As we have extensively discussed, however,
 the presence of $u$-channel singularities makes it inapplicable to
Compton scattering in full generality.

%\vskip .5cm
%for many clarifying discussions on the matter discussed in this work.
%I thank the Theory Groups at Stockholm and Lund Universities,
%CNRS Saclay, Academia Sinica of Taiwan, St. Petersberg's Nucl. Phys.
%%Institute,
%and the Boncompagni-Ludovisi Foundation
%for shelter and hospitality during the first stage of the work.

\newpage

\noindent
{\bf Figure Captions}
\bigskip

 \bigskip

 \noindent
{\bf Fig. 1.} (a) The Compton scattering amplitude;
(b) the 3-particle cut for Compton scattering. One internal line is off shell.
(c) The triangle singularity for the form factor to lowest order.
\smallskip

\noindent
{\bf Fig. 2.} The integration contours for the scalar amplitude
in the $p_1^2$ plane,
which includes the $u$-channel threshold (thick black line) .
\smallskip

\noindent
{\bf Fig. 3.} The $u$-channel cut of the box diagram.
\smallskip

\noindent
{\bf Fig. 4}. (a) The diagram for the gluonic power corrections.
(b) The 3-particle cut for the same diagram;
(c) its $u$ channel cut.
\smallskip

\noindent
{\bf Fig. 5.} (a) Radiative correction to the box diagram in the scalar
amplitude. The line with the encircled "1" denoting a scalar of mass $\mu$.
(b) A cut generated by (a). \newline
(c) Its decomposition as a convolution of
a 2-particle cut in the $s'$ energy (bottom part) with a 2-to-2 on-shell
scattering amplitude. (d) The cut
giving the imaginary contribution generated by (a) and (e) its factorized form.
\smallskip

\noindent
{\bf Fig. 6} (a) A subdiagram of vertex type with a 2-particle cut.
(b) Its factorized form.
(c) The self-energy 2-particle cut. (c) The imaginary contribution
to the spectral density obtained by the insertion of (b).
\smallskip

\noindent
{\bf Fig. 7} The complete OPE of the imaginary parts of the double
spectral densities for Compton scattering.
\smallskip

\noindent
{\bf Fig. 8} (a) A diagram containing a cut of type $t$.
(b) The discontinuity of $t$ type extracted from (a).
(c) A possible (real) cut which is not factorizable.
(d) The same as in (b). Notice that 3 propagators are off-shell.
\smallskip

%%%%%%%%%%%%%%%%%%%%
\end{document}